\documentclass[structabstract]{aa} 

\usepackage{graphicx}
\usepackage{array}
\usepackage{amssymb}
\usepackage{natbib}
\usepackage{float}
\usepackage{color}
\usepackage[utf8]{inputenc}  
\usepackage{amsmath}  
\usepackage{comment}
\usepackage{footnote}
\usepackage{longtable}
\usepackage[caption2]{ccaption}
\usepackage{supertabular}

\bibpunct{(}{)}{;}{a}{,}{,}

\newcommand{\degree}{\ensuremath{^\circ}}
\begin{document}
\title{Magnetic field geometry of an unusual cometary cloud Gal 110-13}

   \subtitle{}

   \author{Neha, S.\inst{1,2}
          \and
          Maheswar, G.\inst{1}
          \and
          Soam, A.\inst{1,2}
          \and
          Chang Won Lee\inst{3,4}
          \and
          Anandmayee Tej\inst{5}
          }

   \institute{Aryabhatta Research Institute of Observational Sciences (ARIES), Nainital 263002, India.\\
              \email{pathakneha.sharma@gmail.com}
   			  \and
              Pt. Ravishankar Shukla University, Raipur, 492010, India.
         	 \and
             Korea Astronomy $\&$ Space Science Institute (KASI), 776 Daedeokdae-ro, Yuseong-gu, Daejeon, Republic of Korea.
             \and
             University of Science Technology, 217 Gajungro, Yuseong-gu, Daejeon 305-333, Korea.
             \and
             Indian Institute of Space Science and Technology (IIST), Trivandrum, India
             }

   \date{Received......, Accepted.....}

 
  \abstract
   {}
   { We carried out optical polarimetry of an isolated cloud, Gal 110-13, to map the plane-of-the-sky magnetic field geometry. The main aim of the study is to understand the most plausible mechanism responsible for the unusual cometary shape of the cloud in the context of its magnetic field geometry.}
   {When unpolarized starlight passes through the intervening interstellar dust grains that are aligned with their short axes parallel to the local magnetic field, it gets linearly polarized. The plane-of-the-sky magnetic field component can therefore be traced by doing polarization measurements of background stars projected on clouds. Because the light in the optical wavelength range is most efficiently polarized by the dust grains typically found in the outer layers of the molecular clouds, optical polarimetry enables us to trace the magnetic field geometry of the outer layers of the clouds.}
   {We made R-band polarization measurements of 207 stars in the direction of Gal 110-13. The distance of Gal 110-13 was determined as $\sim450\pm80$ pc using our polarization and 2MASS near-infrared data. The foreground interstellar contribution was removed from the observed polarization values by observing a number of stars located in the vicinity of Gal 110-13 which has Hipparcos parallax measurements. The plane-of-the-sky magnetic field lines are found to be well ordered and aligned with the elongated structure of Gal 110-13. Using structure function analysis, we estimated the strength of the plane-of-the-sky component of the magnetic field as $\sim25\mu$G.}
   {Based on our results and comparing them with those from simulations, we conclude that compression by the ionization fronts from 10 Lac is the most plausible cause of the comet-like morphology of Gal 110-13 and of the initiation of subsequent star formation.}


	\titlerunning{Magnetic field geometry of Gal 110-13}
   \maketitle
%


\section{Introduction}

Gal 110-13, also known as LBN 534 or DG 191, is an isolated, unusually elongated, comet-shaped molecular cloud located at the galactic coordinate \emph{l} = 110$\degree$ \& \emph{b} = $-13\degree$. The cloud belongs to the Andromeda constellation. First recognized by \citet{1949ApJ...109..540W}, the Gal 110-13 cloud consists of a blue reflection nebula, vdB 158, located at the southern end of the cloud, which is illuminated by a B-type star, HD 222142. Based on the spectroscopic parallaxes determined for HD 222142 and another B-type star, HD 222086, \citet{1969AJ.....74.1021A} derived a distance of 440$\pm$100 pc to the cloud. Using 100$\mu$m IRAS data, \citet{1987ApJ...318..702O} identified and listed 15 isolated clouds that showed cometary or filamentary morphology. The remarkable comet-like morphology of Gal 110-13 makes it one of the most unusual and isolated clouds in the list by \citet{1987ApJ...318..702O}. The cloud major axis, which is $\sim1\degree$ in extent, is oriented at a position angle of $\sim45\degree$ to the east with respect to the north. The cloud has a width of $\sim8^{\prime}$. 

\citet{1992ApJ...397..174O} made a detailed study of Gal 110-13 multiwavelength observations. They found the cloud, especially the tail part, to be highly clumpy in nature. They explored possible mechanisms (such as the interaction of the cloud with ionization fronts or supernova remnants, gravitational instability, hydrodynamics process involving gas stripping, and the cloud-cloud collision) that could explain the unusual shape of the cloud. Of these, according to \citet{1992ApJ...397..174O}, the cloud-cloud collision mechanism was considered to be the most plausible one. Gal 110-13 is explained to have formed as a consequence of collision between two interstellar clouds whose compression zone resulted in the elongated far-IR and $^{12}$CO emission detected from the cloud. However, because Gal 110-13 is roughly pointing toward the Lac OB association, \citet{2007ApJ...657..884L} suggest that the present morphology of the cloud could also be explained by its interaction with supernova blast waves or ionization fronts from the massive members of the Lac OB association. A comet-like morphology of clouds is formed when a supernova explosion shocks a pre-existing spherical cloud and compresses it to form the head, and the blast wave drives the material mechanically away from the supernova to form the tail \citep{1983MNRAS.203..215B}. \citet{1983A&A...117..183R} suggested that the UV radiation from massive stars photoionizes a pre-existing spherical cloud, and shock fronts compress it to form the head. The tail is formed either from the eroded material of the cloud or the pre-existing medium protected from the UV radiation because of the shadowing of the tail by the head.

Several analytical and numerical hydrodynamics studies have been carried out to understand the dynamical behavior of the collision between interstellar clouds \citep{1970ApJ...159..277S, 1970ApJ...159..293S, 1981ApJ...245...72H, 1984ApJ...279..335G, 1985MNRAS.215..125L, 1989ApJ...346..184K, 1992PASJ...44..203H, 1996A&A...308..979K, 1997ApJ...485..254R, 1997ApJ...491..216M, 1998ApJ...497..777K, 2000A&AS..142..165M,  2009A&A...504..437A, 2010MNRAS.405.1431A,2014ApJ...792...63T} and a dense, isolated cloud that is illuminated from one side by a source of ionizing radiation \citep{1989ApJ...346..735B, 1990ApJ...354..529B, 1992IAUS..150..133S, 1994A&A...289..559L, 1995A&A...301..522L, 2001MNRAS.327..788W, 2003MNRAS.338..545K, 2005ApJ...621..803M, 2009ApJ...692..382M, 2010ApJ...723..971G, 2010MNRAS.403..714M, 2012A&A...546A..33T}. In recent years, a number of studies have been conducted to address the role played by the magnetic field in the cloud-cloud collision scenario \citep[e.g., ][]{2001A&A...379.1123M} and in the external ionization-induced evolution of an isolated globule \citep[e.g., ][]{2009MNRAS.398..157H}. In both these processes, the magnetic field seems to alter the results from the hydrodynamical evolution significantly.

Based on 3D-smoothed particle magnetohydrodynamics simulations \citep[3D-SPMHD, ][]{2001A&A...379.1123M}, it was shown that in a cloud-cloud collision scenario the clump formation is only possible in the case where the collision happens between two identical spherical clouds that are traveling parallel to the magnetic field orientation. After the collision, the field lines are found to be chaotic on the cloud scale ($\sim$few pc) though a coherence in the field distribution was found on the clump scale ($\lesssim0.5$ pc). The evolution of globules under the influence of ionizing radiation from an external source was studied using 3D-radiation magnetohydrodynamics simulations \citep{2009MNRAS.398..157H, 2011MNRAS.412.2079M}. The studies show that a strong ($\sim 150-180\mu G$) initially uniform magnetic field perpendicular to the direction of the ionizing radiation remained unchanged, thus altering the dynamical evolution of the globule significantly from the non-magnetic case. In contrast, weak or medium, initially perpendicular magnetic field lines are swept into alignment with the direction of the ionization radiation and the long axis of the globule during the evolution. The magnetic field parallel to the direction of the ionizing radiation, however, are found to remain unchanged producing a broader and snubber globule head. Irrespective of the initial magnetic field orientation, the field lines are found to be ordered well in the external radiation-induced evolution of globules. Such well-ordered magnetic field lines are observed in a number of cometary globules \citep[e.g., ][]{1996MNRAS.279.1191S, 1999MNRAS.308...40B, 2004MNRAS.348...83B, 2011ApJ...743...54T, 2013MNRAS.432.1502S}. Therefore, we expect that the geometry of the magnetic field lines in Gal 110-13 provide useful information on the possible mechanism that might have been responsible for forming the cloud. 

Observations of polarized radiation in the interstellar medium at optical-through-millimeter wavelengths have been attributed to extinction by, and emission from, interstellar dust grains \citep{1949Natur.163..283H, 1951ApJ...114..241H, 1988QJRAS..29..327H}. The polarization measurements in optical \citep[e.g., ][]{1976AJ.....81..958V, 1990ApJ...359..363G, 2008A&A...486L..13A, 2013MNRAS.432.1502S, 2014A&A...569L...1A}, near-infrared \citep[e.g., ][]{1995ApJ...448..748G, 2011ApJ...741...21C, 2011ApJ...734...63S, 2012ApJ...748...18C, 2014ApJ...793..126C, 2014A&A...565A..94B, 2015ApJ...798...60K, 2014A&A...569L...1A} and submillimeter-millimeter wavelengths \citep[e.g., ][]{1998ApJ...502L..75R, 2000ApJS..128..335D, 2010ApJS..186..406D, 2012ApJS..201...13V, 2014ApJS..213...13H, 2014A&A...569L...1A} are used to map the magnetic field geometry of molecular clouds. Optical polarization position angles trace the plane of the sky orientation of the ambient magnetic field at the periphery of the molecular clouds \citep[with $A_{V}$ $\sim$ 1-2 mag; ][]{1995ApJ...448..748G, 1996ASPC...97..325G}. In this work, we present the optical polarization measurements of 207 stars projected on Gal 110-13 to map the projected magnetic field geometry of the cloud. We present the details of the observations and data reduction in section \ref{obs_dr}. In section \ref{res}, we present the results obtained from the observations and in section \ref{dis} discuss the results. We conclude the paper with a summary presented in section \ref{con}.


\section{Observations and data reduction}\label{obs_dr}

The optical linear polarization observations of the region associated with Gal 110-13 were carried out with the ARIES IMaging POLarimeter (AIMPOL), mounted at the f/13 Cassegrain focus of the 104-cm Sampurnanand telescope of ARIES, Nainital, India. It consists of a field lens in combination with the camera lens (85mm, f/1.8); in between them, an achromatic, rotatable half-wave plate (HWP) are used as modulator and a Wollaston prism beam-splitter as analyzer. A detailed description of the instrument and the techniques of the polarization measurements may be found in \cite{2004BASI...32..159R}. We used a standard Johnson R$_\mathrm{KC}$ filter ($\lambda_{\mathrm{eff}}$ = 0.760 $\mu$m) photometric band for the polarimetric observations. The plate scale of the CCD is $1.48^{\prime\prime}$/pixel and the field of view is $\sim 8^{\prime}$ in diameter. The full width at half maximum (FWHM) varies from two to three pixels. The read-out noise and the gain of the CCD are 7.0 e$^{-1}$ and 11.98 e$^{-1}$/ADU, respectively. The log of the observations is presented in Table \ref{table:1}.

\begin{center}
 \begin{table}
	\caption{Log of observations.}\label{table:1}
	\begin{minipage}{\textwidth}   
 \begin{tabular}{ c l } \hline
 Cloud ID & Date of observations (year, month, date) \\ \hline
   Gal 110-13 & 2014 October 19, 20, 21, 22, 23  \\
   			 & 2014 November 17, 18, 19     \\ \hline
 
 \end{tabular}
 
 \end{minipage}
 \end{table}
 \end{center}

We used \emph{Image Reduction \& Analysis Facility (IRAF)} package to perform standard aperture photometry to extract the fluxes of ordinary (I$_{o}$) and extraordinary (I$_{e}$) rays for all the observed sources with a good signal-to-noise ratio.
When the HWP is rotated by an angle $\alpha$, then the electric field vector rotates by an angle 2$\alpha$. A ratio R($\alpha$) has been defined as follows:
\begin{equation}
R(\alpha) = \frac{\frac{I_{e}(\alpha)}{I_{o}(\alpha)}-1}{\frac{I_{e}(\alpha)}{I_{o}(\alpha)}+1} = P \cos(2\theta - 4\alpha)
\end{equation}  

where P is the fraction of the total linearly polarized light, and $\theta$ the polarization angle of the plane of polarization. Since $\alpha$ is the position of the fast axis of the HWP at 0$\degree$, 22.5$\degree$, 45$\degree$ and 67.5$\degree$ corresponding to the four normalized Stokes parameters, respectively, q [R(0$\degree$)], u [R(22.5$\degree$)], q1 [R(45$\degree$)] and u1 [R(67.5$\degree$)]. Because the polarization accuracy is, in principle, limited by the photon statistics, the errors in normalized Stokes parameters $\sigma_{R}(\alpha)$ ($\sigma_{q}$, $\sigma_{u}$, $\sigma_{q1}$ and $\sigma_{u1}$ in percent) are estimated using the expression \citep{1998A&AS..128..369R}

\begin{equation}
\sigma_{R}(\alpha) = \frac{\sqrt{N_{e}+N_{o}+2N_{b}}}{N_{e}+ N_{o}}
\end{equation}

where N$_{o}$ and N$_{e}$ are the number of photons in ordinary and extraordinary rays, respectively, and N$_{\mathrm{b}}$[= (N$_{\mathrm{be}}$+N$_{\mathrm{bo}}$)/2] is the average background photon counts in the vicinity of the extraordinary and ordinary rays.

Standard stars with zero polarization were observed during each run to check for any possible instrumental polarization. The typical instrumental polarization is found to be less than 0.1 $\%$. The instrumental polarization of AIMPOL on the 104-cm Sampurnanand telescope has been monitored since 2004 for various observing programs and found to be stable. Two polarized standard stars HD 236633 and BD$+$59$\degree$389 chosen from \cite{1992AJ....104.1563S} were observed on every observing run to determine the reference direction of the polarizer. In addition another polarized standard star HD 19820 was also observed for two nights. The results obtained for the standard stars are presented in Table \ref{table:2}. The polarization values for the standard stars given in \cite{1992AJ....104.1563S} were obtained with the Kron-Cousins R filter. We corrected for the instrumental polarization from the observed degree of polarization and for the zero point offset using the offset between the standard polarization position angle values obtained by observation and those given in  \cite{1992AJ....104.1563S}.

\begin{table}
	\caption{Polarized standard stars observed in$\ R_{\mathrm{kc}}$ band.}\label{table:2}
	\begin{minipage}{\textwidth}  
 \begin{tabular}{ c c c } \hline
 Date of observations & P $\pm$ $\epsilon_{\mathrm{P}}$  & $\theta$ $\pm$ ${\epsilon_{\theta}}$ \\ 
          &            $(\%)$                &              $(\degree)$             \\ \hline

\multicolumn{3}{c}{\bf HD236633}\\
 Standard value$^{\dagger}$&5.38$\pm$0.02&93.04$\pm0.15$\\
 19 Oct 2014 & 4.8 $\pm$ 0.2 &  95 $\pm$ 1 \\
 20 Oct 2014 & 4.9 $\pm$ 0.1 &  94 $\pm$ 1 \\
 21 Oct 2014 & 4.7 $\pm$ 0.2 &  93 $\pm$ 1 \\
 22 Oct 2014 & 5.1 $\pm$ 0.1 &  93 $\pm$ 1 \\
 23 Oct 2014 & 5.0 $\pm$ 0.1 &  92 $\pm$ 1 \\ 
\multicolumn{3}{c} {\bf BD+59$\degree$389}\\
  Standard value$^{\dagger}$&6.43$\pm$0.02&98.1$\pm0.1$\\
  19 Oct 2014 & 6.0 $\pm$ 0.1 &  97 $\pm$ 1 \\
  20 Oct 2014 & 5.8 $\pm$ 0.1 &  99 $\pm$ 1 \\
  21 Oct 2014 & 6.2 $\pm$ 0.1 &  98 $\pm$ 1 \\
  22 Oct 2014 & 5.8 $\pm$ 0.2 &  97 $\pm$ 1 \\
  23 Oct 2014 & 6.2 $\pm$ 0.1 &  98 $\pm$ 1 \\ 
  18 Nov 2014 & 6.4 $\pm$ 0.1 &  98 $\pm$ 1 \\
  19 Nov 2014 & 6.3 $\pm$ 0.1 &  98 $\pm$ 1 \\  
\multicolumn{3}{c} {\bf HD19820}\\
  Standard value$^{\dagger}$&4.53$\pm$0.03&114.46$\pm0.16$\\  
  18 Nov 2014 & 4.5 $\pm$ 0.1 &  114 $\pm$ 1 \\
  19 Nov 2014 & 4.5 $\pm$ 0.1 &  115 $\pm$ 1 \\ \hline
 \end{tabular}

$^{\dagger}$Standard values in the R$-$band taken from \cite{1992AJ....104.1563S}.
 \end{minipage}
 \end{table}

\section{RESULTS} \label{res}

\begin{figure}[h!]
\centering
\resizebox{8.5cm}{15.0cm}{\includegraphics{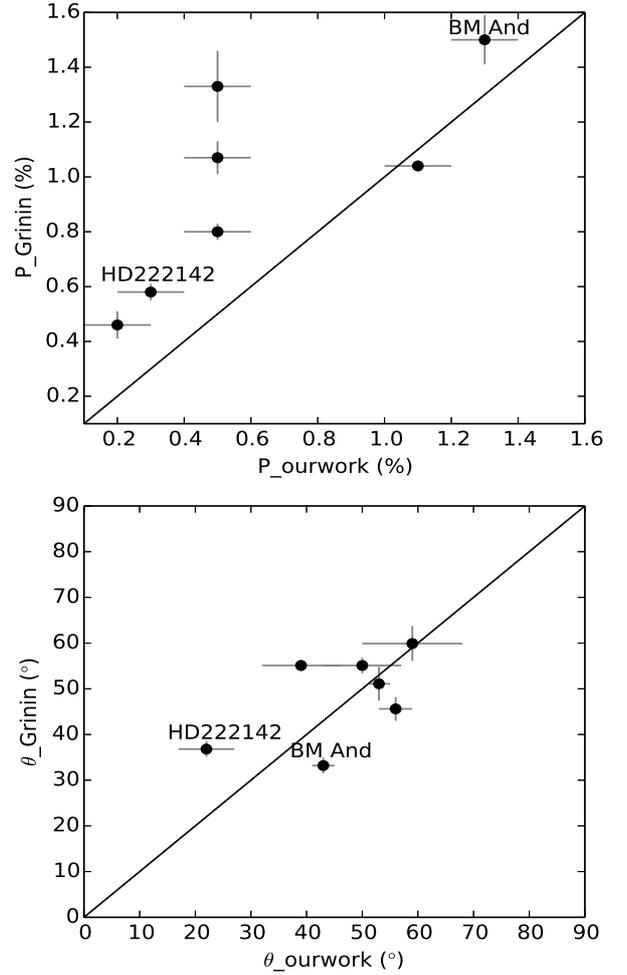}}
\caption{Comparison of the degree of polarization (upper panel) and the position angles (lower panel) of the seven stars found in common with those observed by Grinin et al (1995).}\label{fig:grinin}
\end{figure}

The results of our R-band polarimetry of 207 stars toward Gal 110-13 are presented in Table 3. We tabulated the results of only those sources that have the ratio of the degree of polarization (P$\%$) and error in the degree of polarization ($\sigma_{\mathrm{P}}$), P/$\sigma_{\mathrm{P}}$ $\geq$ 2. The columns of the Table 3 show the star identification number in increasing order of their right ascension, the declination, measured P ($\%$) and polarization position angles ($\theta_P$ in degrees). The mean value of P is found to be 1.5\%. The mean value of $\theta_P$ obtained from a Gaussian fit to the distribution is found to be 45$\degree$. The standard deviation in P and $\theta_P$ is found to be 0.8\% and 12$\degree$, respectively. A total of ten stars were observed by \citet{1995A&AS..112..457G} toward the Gal 110-13 region. Of these, seven sources are found to be in common. In Fig. \ref{fig:grinin}, we compare our results with those from the \citet{1995A&AS..112..457G} obtained in R filter. The effective wavelength of the R filter used is not mentioned in the paper. The degree of polarization obtained by \citet{1995A&AS..112..457G} is found to be systematically higher. Two sources, BM And and HD 222142 are identified and labeled. BM And is classified as a classical T Tauri-type star \citep{1988cels.book.....H}. This star is known for its photometric and polarimetric variability \citep{1995A&AS..112..457G}. HD 222142 which is associated with the reflection nebulosity could also be a variable source.

\begin{table}
\begin{minipage}{80mm}
\caption{Polarization results of 207 stars (with P/$\sigma_{\mathrm{P}}$ $\ge$ 2) observed in the direction of Gal 110-13.}\label{tab:res70}
\begin{tabular}{ccccr}  \hline
Star & $\alpha$ (J2000)  & $\delta$ (J2000)  & P $\pm$ $\epsilon_P$ & $\theta$ $\pm$ $\epsilon_{\theta}$  \\ 
 Id  &   ($\degree$) &    ($\degree$)&    (\%)           & ($\degree$) \\\hline
1	 & 	354.016876	 & 	48.341923	 & 	1.4 $\pm$ 0.2	 & 	47 $\pm$ 4 \\ 
2	 & 	354.032440	 & 	48.305222	 & 	0.4 $\pm$ 0.2	 & 	67 $\pm$ 10 \\ 
3	 & 	354.068573	 & 	48.540981	 & 	1.0 $\pm$ 0.4	 & 	28 $\pm$ 11 \\ 
4	 & 	354.072418	 & 	48.359886	 & 	0.9 $\pm$ 0.1	 & 	38 $\pm$ 2 \\ 
5	 & 	354.075134	 & 	48.556656	 & 	1.1 $\pm$ 0.1	 & 	42 $\pm$ 3 \\ 
6	 & 	354.079865	 & 	48.484379	 & 	1.4 $\pm$ 0.5	 & 	56 $\pm$ 9 \\ 
7	 & 	354.085388	 & 	48.487614	 & 	1.3 $\pm$ 0.3	 & 	35 $\pm$ 7 \\ 
8	 & 	354.098969	 & 	48.516445	 & 	1.1 $\pm$ 0.2	 & 	43 $\pm$ 5 \\ 
9	 & 	354.099731	 & 	48.310612	 & 	1.3 $\pm$ 0.3	 & 	38 $\pm$ 7 \\ 
10	 & 	354.099976	 & 	48.374210	 & 	1.5 $\pm$ 0.4	 & 	15 $\pm$ 7 \\ 
\multicolumn{5}{c}{}\\ 
11	 & 	354.103058	 & 	48.378044	 & 	1.4 $\pm$ 0.5	 & 	43 $\pm$ 9 \\ 
12	 & 	354.111816	 & 	48.296593	 & 	1.2 $\pm$ 0.2	 & 	27 $\pm$ 5 \\ 
13	 & 	354.116089	 & 	48.476658	 & 	0.6 $\pm$ 0.1	 & 	64 $\pm$ 3 \\ 
14	 & 	354.117523	 & 	48.295223	 & 	0.6 $\pm$ 0.1	 & 	29 $\pm$ 4 \\ 
15	 & 	354.117706	 & 	48.341057	 & 	1.8 $\pm$ 0.2	 & 	20 $\pm$ 3 \\ 
16	 & 	354.129303	 & 	48.520378	 & 	1.6 $\pm$ 0.4	 & 	30 $\pm$ 7 \\ 
17	 & 	354.135895	 & 	48.517021	 & 	0.6 $\pm$ 0.1	 & 	46 $\pm$ 6 \\ 
18	 & 	354.136749	 & 	48.503525	 & 	0.9 $\pm$ 0.2	 & 	87 $\pm$ 5 \\ 
19	 & 	354.139648	 & 	48.536335	 & 	1.1 $\pm$ 0.2	 & 	38 $\pm$ 6 \\ 
20	 & 	354.140961	 & 	48.547497	 & 	1.4 $\pm$ 0.4	 & 	53 $\pm$ 7 \\ 
\multicolumn{5}{c}{}\\ 
21	 & 	354.143494	 & 	48.328995	 & 	2.6 $\pm$ 0.8	 & 	37 $\pm$ 9 \\ 
22	 & 	354.148163	 & 	48.526020	 & 	1.0 $\pm$ 0.2	 & 	33 $\pm$ 4 \\ 
23	 & 	354.154785	 & 	48.514671	 & 	1.7 $\pm$ 0.6	 & 	58 $\pm$ 9 \\ 
24	 & 	354.156464	 & 	48.445515	 & 	1.8 $\pm$ 0.1	 & 	35 $\pm$ 1 \\ 
25	 & 	354.157623	 & 	48.317368	 & 	1.6 $\pm$ 0.4	 & 	176 $\pm$ 7 \\ 
26	 & 	354.158722	 & 	48.312260	 & 	2.6 $\pm$ 0.6	 & 	7 $\pm$ 7 \\ 
27	 & 	354.159302	 & 	48.546272	 & 	1.4 $\pm$ 0.3	 & 	41 $\pm$ 7 \\ 
28	 & 	354.173462	 & 	48.438557	 & 	2.6 $\pm$ 0.8	 & 	36 $\pm$ 9 \\ 
29	 & 	354.177917	 & 	48.435986	 & 	2.0 $\pm$ 0.4	 & 	39 $\pm$ 6 \\ 
30	 & 	354.180389	 & 	48.325775	 & 	2.1 $\pm$ 0.4	 & 	27 $\pm$ 5 \\ 
\multicolumn{5}{c}{}\\ 
31	 & 	354.190399	 & 	48.526745	 & 	0.5 $\pm$ 0.2	 & 	49 $\pm$ 10 \\ 
32	 & 	354.191528	 & 	48.328285	 & 	2.5 $\pm$ 0.7	 & 	25 $\pm$ 7 \\ 
33	 & 	354.191803	 & 	48.333572	 & 	2.5 $\pm$ 1.3	 & 	35 $\pm$ 14 \\ 
34	 & 	354.200775	 & 	48.459743	 & 	2.5 $\pm$ 0.4	 & 	36 $\pm$ 5 \\ 
35	 & 	354.204987	 & 	48.445213	 & 	2.5 $\pm$ 0.5	 & 	26 $\pm$ 6 \\ 
36	 & 	354.215057	 & 	48.441216	 & 	1.8 $\pm$ 0.2	 & 	32 $\pm$ 2 \\ 
37	 & 	354.218445	 & 	48.506569	 & 	1.3 $\pm$ 0.2	 & 	47 $\pm$ 4 \\ 
38	 & 	354.220947	 & 	48.490379	 & 	2.1 $\pm$ 0.5	 & 	45 $\pm$ 6 \\ 
39	 & 	354.223663	 & 	48.278530	 & 	1.6 $\pm$ 0.4	 & 	37 $\pm$ 7 \\ 
40	 & 	354.226593	 & 	48.301861	 & 	1.9 $\pm$ 0.6	 & 	28 $\pm$ 9 \\ 
\multicolumn{5}{c}{}\\ 
41	 & 	354.229187	 & 	48.421490	 & 	0.6 $\pm$ 0.2	 & 	50 $\pm$ 10 \\ 
42	 & 	354.231934	 & 	48.268093	 & 	1.3 $\pm$ 0.3	 & 	40 $\pm$ 5 \\ 
43	 & 	354.237518	 & 	48.517609	 & 	1.1 $\pm$ 0.1	 & 	44 $\pm$ 4 \\ 
44	 & 	354.238739	 & 	48.488358	 & 	2.4 $\pm$ 1.1	 & 	34 $\pm$ 12 \\ 
45	 & 	354.240753	 & 	48.279388	 & 	1.6 $\pm$ 0.5	 & 	27 $\pm$ 8 \\ 
46	 & 	354.244598	 & 	48.493759	 & 	2.6 $\pm$ 0.2	 & 	42 $\pm$ 2 \\ 
47	 & 	354.246918	 & 	48.412663	 & 	1.1 $\pm$ 0.4	 & 	112 $\pm$ 11 \\ 
48	 & 	354.247803	 & 	48.273891	 & 	2.1 $\pm$ 0.7	 & 	43 $\pm$ 9 \\ 
49	 & 	354.251709	 & 	48.380104	 & 	0.6 $\pm$ 0.2	 & 	108 $\pm$ 9 \\ 
50	 & 	354.256348	 & 	48.274315	 & 	1.7 $\pm$ 0.2	 & 	34 $\pm$ 4 \\ 
\multicolumn{5}{c}{}\\ 
51$^{a}$	 & 	354.257130	 & 	48.576075	 & 	0.5 $\pm$ 0.1	 & 	50 $\pm$ 7 \\ 
52	 & 	354.259857	 & 	48.486195	 & 	1.7 $\pm$ 0.4	 & 	44 $\pm$ 6 \\ 
53	 & 	354.266022	 & 	48.315334	 & 	1.8 $\pm$ 0.7	 & 	44 $\pm$ 11 \\ 
54	 & 	354.275604	 & 	48.546604	 & 	1.4 $\pm$ 0.1	 & 	48 $\pm$ 3 \\ 
55	 & 	354.283386	 & 	48.318745	 & 	2.0 $\pm$ 0.5	 & 	61 $\pm$ 7 \\ 
56	 & 	354.284576	 & 	48.513348	 & 	3.4 $\pm$ 0.5	 & 	50 $\pm$ 4 \\ 
57	 & 	354.290161	 & 	48.430916	 & 	1.4 $\pm$ 0.6	 & 	124 $\pm$ 11 \\ 
58	 & 	354.292023	 & 	48.480244	 & 	1.4 $\pm$ 0.6	 & 	59 $\pm$ 12 \\ 
59	 & 	354.293152	 & 	48.279785	 & 	2.5 $\pm$ 0.4	 & 	32 $\pm$ 4 \\ 
60	 & 	354.302094	 & 	48.519394	 & 	2.3 $\pm$ 0.8	 & 	45 $\pm$ 10 \\ 
\end{tabular}
Spectral type $- ^{a}$B9V
\end{minipage}
\end{table}
\begin{table}
\centering
\begin{minipage}{80mm}
\contcaption{Polarization results of 207 stars (with P/$\sigma_{\mathrm{P}}$ $\ge$ 2) observed in the direction of Gal 110-13.}
\begin{tabular}{ccccr}  \hline
Star  & $\alpha$ (J2000)  & $\delta$ (J2000)  & P $\pm$ $\epsilon_P$ & $\theta$ $\pm$ $\epsilon_{\theta}$  \\ 
Id  &($\degree$)&($\degree$)& (\%) &($\degree$) \\\hline 
61	 & 	354.302155	 & 	48.347641	 & 	3.3 $\pm$ 1.3	 & 	42 $\pm$ 11 \\ 
62	 & 	354.307617	 & 	48.336323	 & 	1.4 $\pm$ 0.1	 & 	36 $\pm$ 3 \\ 
63	 & 	354.308289	 & 	48.558563	 & 	0.9 $\pm$ 0.4	 & 	109 $\pm$ 13 \\ 
64	 & 	354.311798	 & 	48.550259	 & 	2.2 $\pm$ 0.6	 & 	34 $\pm$ 7 \\ 
65	 & 	354.311798	 & 	48.503056	 & 	2.5 $\pm$ 1.0	 & 	58 $\pm$ 12 \\ 
66	 & 	354.315887	 & 	48.285450	 & 	2.0 $\pm$ 0.6	 & 	41 $\pm$ 9 \\ 
67	 & 	354.321716	 & 	48.522610	 & 	1.9 $\pm$ 0.3	 & 	12 $\pm$ 5 \\ 
68	 & 	354.322418	 & 	48.295036	 & 	1.1 $\pm$ 0.5	 & 	51 $\pm$ 12 \\ 
69$^{b}$	 & 	354.325780	 & 	48.482020	 & 	0.5 $\pm$ 0.1	 & 	39 $\pm$ 7 \\ 
70	 & 	354.326782	 & 	48.313019	 & 	0.6 $\pm$ 0.2	 & 	50 $\pm$ 9 \\ 
\multicolumn{5}{c}{}\\ 
71	 & 	354.330017	 & 	48.644291	 & 	1.6 $\pm$ 0.8	 & 	47 $\pm$ 13 \\ 
72	 & 	354.341217	 & 	48.332596	 & 	2.0 $\pm$ 0.7	 & 	42 $\pm$ 10 \\ 
73	 & 	354.348022	 & 	48.636242	 & 	0.4 $\pm$ 0.1	 & 	56 $\pm$ 6 \\ 
74	 & 	354.348816	 & 	48.327698	 & 	2.1 $\pm$ 0.6	 & 	51 $\pm$ 7 \\ 
75	 & 	354.361969	 & 	48.300316	 & 	2.6 $\pm$ 0.1	 & 	34 $\pm$ 1 \\ 
76	 & 	354.367340	 & 	48.624557	 & 	0.3 $\pm$ 0.1	 & 	72 $\pm$ 4 \\ 
77$^{c}$	 & 	354.369950	 & 	48.541050	 & 	1.1 $\pm$ 0.1	 & 	53 $\pm$ 2 \\ 
78	 & 	354.375488	 & 	48.649315	 & 	1.2 $\pm$ 0.4	 & 	27 $\pm$ 10 \\ 
79	 & 	354.379456	 & 	48.622993	 & 	4.3 $\pm$ 1.1	 & 	40 $\pm$ 8 \\ 
80	 & 	354.381104	 & 	48.488827	 & 	1.5 $\pm$ 0.2	 & 	41 $\pm$ 3 \\ 
\multicolumn{5}{c}{}\\ 
81	 & 	354.392853	 & 	48.483425	 & 	1.9 $\pm$ 0.8	 & 	52 $\pm$ 12 \\ 
82	 & 	354.398804	 & 	48.693077	 & 	1.1 $\pm$ 0.4	 & 	54 $\pm$ 9 \\ 
83$^{d}$	 & 	354.410339	 & 	48.403324	 & 	1.3 $\pm$ 0.1	 & 	43 $\pm$ 2 \\ 
84	 & 	354.412384	 & 	48.644550	 & 	2.0 $\pm$ 0.4	 & 	49 $\pm$ 5 \\ 
85	 & 	354.414276	 & 	48.609211	 & 	0.9 $\pm$ 0.3	 & 	58 $\pm$ 8 \\ 
86	 & 	354.414337	 & 	48.459354	 & 	2.1 $\pm$ 0.8	 & 	40 $\pm$ 11 \\ 
87	 & 	354.418091	 & 	48.414818	 & 	3.3 $\pm$ 0.9	 & 	36 $\pm$ 8 \\ 
88	 & 	354.423279	 & 	48.423710	 & 	2.2 $\pm$ 1.0	 & 	42 $\pm$ 13 \\ 
89	 & 	354.424652	 & 	48.447205	 & 	1.5 $\pm$ 0.1	 & 	29 $\pm$ 2 \\ 
90	 & 	354.431061	 & 	48.658772	 & 	0.8 $\pm$ 0.1	 & 	40 $\pm$ 2 \\ 
\multicolumn{5}{c}{}\\ 
91	 & 	354.444489	 & 	48.597054	 & 	0.8 $\pm$ 0.2	 & 	40 $\pm$ 6 \\ 
92	 & 	354.448303	 & 	48.640835	 & 	2.7 $\pm$ 0.4	 & 	51 $\pm$ 4 \\ 
93	 & 	354.455750	 & 	48.346878	 & 	1.5 $\pm$ 0.4	 & 	36 $\pm$ 7 \\ 
94	 & 	354.458618	 & 	48.463482	 & 	2.0 $\pm$ 0.2	 & 	33 $\pm$ 3 \\ 
95	 & 	354.459839	 & 	48.356136	 & 	1.2 $\pm$ 0.5	 & 	42 $\pm$ 11 \\ 
96	 & 	354.463837	 & 	48.425594	 & 	1.7 $\pm$ 0.6	 & 	7 $\pm$ 9 \\ 
97$^{e}$ & 	354.465057	 & 	48.496559	 & 	0.3 $\pm$ 0.1	 & 	22 $\pm$ 5 \\ 
98	 & 	354.470642	 & 	48.384491	 & 	1.9 $\pm$ 0.3	 & 	44 $\pm$ 5 \\ 
99	 & 	354.470795	 & 	48.358414	 & 	1.3 $\pm$ 0.3	 & 	32 $\pm$ 5 \\ 
100	 & 	354.472382	 & 	48.676861	 & 	0.8 $\pm$ 0.1	 & 	49 $\pm$ 3 \\ 
\multicolumn{5}{c}{}\\ 
101	 & 	354.480499	 & 	48.396702	 & 	1.5 $\pm$ 0.4	 & 	42 $\pm$ 8 \\ 
102	 & 	354.484161	 & 	48.466385	 & 	2.0 $\pm$ 0.2	 & 	31 $\pm$ 3 \\ 
103	 & 	354.485382	 & 	48.656799	 & 	1.8 $\pm$ 0.6	 & 	42 $\pm$ 9 \\ 
104	 & 	354.493652	 & 	48.409611	 & 	1.7 $\pm$ 0.8	 & 	19 $\pm$ 13 \\ 
105	 & 	354.493713	 & 	48.406029	 & 	2.8 $\pm$ 1.0	 & 	32 $\pm$ 10 \\ 
106	 & 	354.507019	 & 	48.827187	 & 	1.2 $\pm$ 0.4	 & 	58 $\pm$ 9 \\ 
107	 & 	354.520844	 & 	48.766190	 & 	0.7 $\pm$ 0.3	 & 	44 $\pm$ 10 \\ 
108	 & 	354.522003	 & 	48.368587	 & 	0.8 $\pm$ 0.3	 & 	48 $\pm$ 10 \\ 
109	 & 	354.522949	 & 	48.801640	 & 	1.7 $\pm$ 0.4	 & 	11 $\pm$ 7 \\ 
110	 & 	354.523499	 & 	48.431416	 & 	3.0 $\pm$ 1.4	 & 	31 $\pm$ 14 \\ 
\multicolumn{5}{c}{}\\ 
111	 & 	354.530304	 & 	48.823254	 & 	1.7 $\pm$ 0.7	 & 	36 $\pm$ 12 \\ 
112	 & 	354.530884	 & 	48.674706	 & 	0.6 $\pm$ 0.2	 & 	63 $\pm$ 7 \\ 
113	 & 	354.532410	 & 	48.697784	 & 	0.8 $\pm$ 0.3	 & 	36 $\pm$ 10 \\ 
114	 & 	354.532928	 & 	48.668297	 & 	0.9 $\pm$ 0.4	 & 	30 $\pm$ 12 \\ 
115	 & 	354.532990	 & 	48.755928	 & 	0.7 $\pm$ 0.1	 & 	50 $\pm$ 2 \\ 
116	 & 	354.533234	 & 	48.391396	 & 	0.9 $\pm$ 0.4	 & 	42 $\pm$ 13 \\ 
117	 & 	354.545898	 & 	48.671093	 & 	2.5 $\pm$ 0.4	 & 	23 $\pm$ 5 \\ 
118	 & 	354.547485	 & 	48.427036	 & 	0.9 $\pm$ 0.4	 & 	45 $\pm$ 12 \\ 
119	 & 	354.551697	 & 	48.689178	 & 	1.3 $\pm$ 0.5	 & 	36 $\pm$ 9 \\ 
120	 & 	354.555420	 & 	48.491554	 & 	1.4 $\pm$ 0.4	 & 	45 $\pm$ 7 \\ 
\end{tabular}
Spectral type $- ^{b}$B9, $^{c}$K0, $^{d}$K5Ve and $^{e}$B8V
\end{minipage}
\end{table}
\begin{table}
\centering
\begin{minipage}{80mm}
\contcaption{Polarization results of 207 stars (with P/$\sigma_{\mathrm{P}}$ $\ge$ 2) observed in the direction of Gal 110-13.}
\begin{tabular}{ccccr}  \hline
Star  & $\alpha$ (J2000)  & $\delta$ (J2000)  & P $\pm$ $\epsilon_P$ & $\theta$ $\pm$ $\epsilon_{\theta}$  \\ 
Id  &($\degree$)&($\degree$)& (\%) &($\degree$) \\\hline 
121	 & 	354.556519	 & 	48.685085	 & 	2.5 $\pm$ 0.7	 & 	24 $\pm$ 8 \\ 
122	 & 	354.558350	 & 	48.665985	 & 	1.6 $\pm$ 0.2	 & 	28 $\pm$ 4 \\ 
123	 & 	354.559753	 & 	48.482159	 & 	3.6 $\pm$ 0.3	 & 	23 $\pm$ 3 \\ 
124	 & 	354.565643	 & 	48.512787	 & 	1.9 $\pm$ 0.7	 & 	25 $\pm$ 10 \\ 
125	 & 	354.569305	 & 	48.527279	 & 	1.6 $\pm$ 0.3	 & 	32 $\pm$ 6 \\ 
126	 & 	354.574219	 & 	48.434845	 & 	1.9 $\pm$ 0.2	 & 	27 $\pm$ 3 \\ 
127	 & 	354.574646	 & 	48.835133	 & 	3.3 $\pm$ 0.6	 & 	36 $\pm$ 5 \\ 
128	 & 	354.574707	 & 	48.439724	 & 	1.3 $\pm$ 0.5	 & 	92 $\pm$ 10 \\ 
129	 & 	354.581543	 & 	48.785568	 & 	1.1 $\pm$ 0.4	 & 	75 $\pm$ 9 \\ 
130	 & 	354.585205	 & 	48.801201	 & 	0.7 $\pm$ 0.1	 & 	56 $\pm$ 4 \\ 
\multicolumn{5}{c}{}\\ 
131	 & 	354.587128	 & 	48.512829	 & 	1.4 $\pm$ 0.7	 & 	57 $\pm$ 12 \\ 
132	 & 	354.592194	 & 	48.680164	 & 	2.1 $\pm$ 0.2	 & 	35 $\pm$ 3 \\ 
133	 & 	354.592865	 & 	48.482475	 & 	0.9 $\pm$ 0.5	 & 	55 $\pm$ 13 \\ 
134	 & 	354.596863	 & 	48.801567	 & 	1.1 $\pm$ 0.5	 & 	122 $\pm$ 11 \\ 
135	 & 	354.612000	 & 	48.685116	 & 	2.3 $\pm$ 0.1	 & 	49 $\pm$ 1 \\ 
136	 & 	354.614075	 & 	48.709793	 & 	1.8 $\pm$ 0.4	 & 	49 $\pm$ 6 \\ 
137	 & 	354.614838	 & 	48.700356	 & 	1.7 $\pm$ 0.2	 & 	58 $\pm$ 4 \\ 
138	 & 	354.618591	 & 	48.808205	 & 	0.8 $\pm$ 0.2	 & 	62 $\pm$ 5 \\ 
139	 & 	354.624420	 & 	48.483952	 & 	1.3 $\pm$ 0.2	 & 	26 $\pm$ 4 \\ 
140	 & 	354.625305	 & 	48.476353	 & 	0.9 $\pm$ 0.4	 & 	45 $\pm$ 12 \\ 
\multicolumn{5}{c}{}\\ 
141	 & 	354.631622	 & 	48.656868	 & 	1.6 $\pm$ 0.2	 & 	51 $\pm$ 4 \\ 
142	 & 	354.638000	 & 	48.532352	 & 	2.0 $\pm$ 0.8	 & 	41 $\pm$ 11 \\ 
143	 & 	354.645996	 & 	48.463760	 & 	1.3 $\pm$ 0.1	 & 	41 $\pm$ 3 \\ 
144	 & 	354.654449	 & 	48.780697	 & 	1.0 $\pm$ 0.4	 & 	43 $\pm$ 10 \\ 
145	 & 	354.654846	 & 	48.526394	 & 	1.3 $\pm$ 0.3	 & 	36 $\pm$ 6 \\ 
146	 & 	354.655243	 & 	48.671700	 & 	1.5 $\pm$ 0.3	 & 	33 $\pm$ 5 \\ 
147	 & 	354.657532	 & 	48.694839	 & 	2.3 $\pm$ 0.5	 & 	47 $\pm$ 6 \\ 
148	 & 	354.658936	 & 	48.685604	 & 	2.1 $\pm$ 0.5	 & 	59 $\pm$ 7 \\ 
149	 & 	354.660492	 & 	48.616344	 & 	1.9 $\pm$ 0.3	 & 	54 $\pm$ 5 \\ 
150	 & 	354.661499	 & 	48.579056	 & 	1.9 $\pm$ 0.3	 & 	46 $\pm$ 5 \\ 
\multicolumn{5}{c}{}\\ 
151	 & 	354.663666	 & 	48.602970	 & 	2.1 $\pm$ 0.5	 & 	40 $\pm$ 6 \\ 
152	 & 	354.665466	 & 	48.663883	 & 	2.4 $\pm$ 0.5	 & 	44 $\pm$ 5 \\ 
153	 & 	354.675201	 & 	48.589035	 & 	1.9 $\pm$ 0.9	 & 	27 $\pm$ 13 \\ 
154	 & 	354.678528	 & 	48.598816	 & 	2.2 $\pm$ 0.8	 & 	27 $\pm$ 10 \\ 
155	 & 	354.679199	 & 	48.572227	 & 	1.6 $\pm$ 0.3	 & 	55 $\pm$ 4 \\ 
156	 & 	354.690430	 & 	48.731926	 & 	1.3 $\pm$ 0.3	 & 	55 $\pm$ 5 \\ 
157	 & 	354.698334	 & 	48.711510	 & 	1.4 $\pm$ 0.2	 & 	53 $\pm$ 4 \\ 
158	 & 	354.698883	 & 	48.544685	 & 	2.9 $\pm$ 0.7	 & 	37 $\pm$ 6 \\ 
159	 & 	354.702789	 & 	48.743034	 & 	1.4 $\pm$ 0.4	 & 	53 $\pm$ 7 \\ 
160	 & 	354.703491	 & 	48.705238	 & 	0.2 $\pm$ 0.1	 & 	54 $\pm$ 9 \\ 
\multicolumn{5}{c}{}\\ 
161	 & 	354.708557	 & 	48.528770	 & 	0.5 $\pm$ 0.1	 & 	56 $\pm$ 3 \\ 
162	 & 	354.710144	 & 	48.629665	 & 	2.0 $\pm$ 0.9	 & 	32 $\pm$ 13 \\ 
163	 & 	354.718628	 & 	48.707359	 & 	1.9 $\pm$ 0.5	 & 	44 $\pm$ 7 \\ 
164	 & 	354.719940	 & 	48.638454	 & 	1.1 $\pm$ 0.2	 & 	43 $\pm$ 6 \\ 
165	 & 	354.721954	 & 	48.693413	 & 	2.1 $\pm$ 0.2	 & 	48 $\pm$ 2 \\ 
166	 & 	354.728241	 & 	48.739994	 & 	2.1 $\pm$ 0.3	 & 	41 $\pm$ 5 \\ 
167	 & 	354.731537	 & 	48.579395	 & 	0.9 $\pm$ 0.3	 & 	75 $\pm$ 7 \\ 
168	 & 	354.736298	 & 	48.608452	 & 	1.2 $\pm$ 0.3	 & 	44 $\pm$ 6 \\ 
169	 & 	354.736877	 & 	48.599663	 & 	1.2 $\pm$ 0.2	 & 	40 $\pm$ 4 \\ 
170	 & 	354.737976	 & 	48.555195	 & 	2.5 $\pm$ 0.4	 & 	2 $\pm$ 5 \\ 
\multicolumn{5}{c}{}\\ 
171	 & 	354.738403	 & 	48.750927	 & 	1.3 $\pm$ 0.2	 & 	34 $\pm$ 5 \\ 
172	 & 	354.742371	 & 	48.747238	 & 	2.4 $\pm$ 0.4	 & 	49 $\pm$ 4 \\ 
173	 & 	354.744263	 & 	48.703583	 & 	0.7 $\pm$ 0.3	 & 	23 $\pm$ 12 \\ 
174	 & 	354.748077	 & 	48.755611	 & 	1.9 $\pm$ 0.3	 & 	36 $\pm$ 5 \\ 
175	 & 	354.753540	 & 	48.765472	 & 	1.0 $\pm$ 0.4	 & 	41 $\pm$ 9 \\ 
176	 & 	354.755005	 & 	48.547642	 & 	2.3 $\pm$ 0.7	 & 	78 $\pm$ 8 \\ 
177	 & 	354.762024	 & 	48.537395	 & 	1.9 $\pm$ 0.7	 & 	24 $\pm$ 9 \\ 
178	 & 	354.764313	 & 	48.556896	 & 	0.8 $\pm$ 0.1	 & 	46 $\pm$ 3 \\ 
179	 & 	354.767578	 & 	48.744648	 & 	1.3 $\pm$ 0.1	 & 	40 $\pm$ 3 \\ 
180	 & 	354.772217	 & 	48.712940	 & 	1.8 $\pm$ 0.5	 & 	44 $\pm$ 8 \\ 
\end{tabular}
\end{minipage}
\end{table}
\begin{table}[h!]
\centering
\begin{minipage}{80mm}
\contcaption{Polarization results of 207 stars (with P/$\sigma_{\mathrm{P}}$ $\ge$ 2) observed in the direction of Gal 110-13.}
\begin{tabular}{ccccr}  \hline
Star  & $\alpha$ (J2000)  & $\delta$ (J2000)  & P $\pm$ $\epsilon_P$ & $\theta$ $\pm$ $\epsilon_{\theta}$  \\ 
Id  &($\degree$)&($\degree$)& (\%) &($\degree$) \\\hline 
181	 & 	354.775513	 & 	48.696392	 & 	1.4 $\pm$ 0.4	 & 	56 $\pm$ 7 \\ 
182	 & 	354.776031	 & 	48.699131	 & 	1.9 $\pm$ 0.7	 & 	57 $\pm$ 10 \\ 
183	 & 	354.778900	 & 	48.550201	 & 	1.2 $\pm$ 0.4	 & 	78 $\pm$ 8 \\ 
184	 & 	354.780609	 & 	48.574768	 & 	0.5 $\pm$ 0.2	 & 	46 $\pm$ 9 \\ 
185	 & 	354.808685	 & 	48.756504	 & 	2.0 $\pm$ 0.5	 & 	33 $\pm$ 7 \\ 
186	 & 	354.822113	 & 	48.695560	 & 	1.0 $\pm$ 0.2	 & 	43 $\pm$ 4 \\ 
187	 & 	354.831482	 & 	48.714211	 & 	1.1 $\pm$ 0.3	 & 	52 $\pm$ 7 \\ 
188	 & 	354.839325	 & 	48.639771	 & 	0.9 $\pm$ 0.4	 & 	46 $\pm$ 11 \\ 
189	 & 	354.846069	 & 	48.701302	 & 	1.5 $\pm$ 0.4	 & 	57 $\pm$ 8 \\ 
190	 & 	354.854523	 & 	48.697876	 & 	1.0 $\pm$ 0.1	 & 	52 $\pm$ 3 \\ 
\multicolumn{5}{c}{}\\ 
191	 & 	354.857178	 & 	48.737354	 & 	1.2 $\pm$ 0.2	 & 	47 $\pm$ 4 \\ 
192	 & 	354.866943	 & 	48.711109	 & 	1.2 $\pm$ 0.1	 & 	50 $\pm$ 4 \\ 
193	 & 	354.878571	 & 	48.625912	 & 	0.3 $\pm$ 0.1	 & 	19 $\pm$ 7 \\ 
194	 & 	354.883179	 & 	48.705502	 & 	1.2 $\pm$ 0.1	 & 	48 $\pm$ 1 \\ 
195	 & 	354.883240	 & 	48.643642	 & 	0.2 $\pm$ 0.1	 & 	59 $\pm$ 8 \\ 
196	 & 	354.892883	 & 	48.658283	 & 	0.9 $\pm$ 0.4	 & 	50 $\pm$ 11 \\ 
197	 & 	354.906006	 & 	48.671249	 & 	0.9 $\pm$ 0.4	 & 	44 $\pm$ 12 \\ 
198	 & 	354.906372	 & 	48.692219	 & 	1.1 $\pm$ 0.3	 & 	63 $\pm$ 6 \\ 
199	 & 	354.911652	 & 	48.606640	 & 	0.5 $\pm$ 0.1	 & 	53 $\pm$ 4 \\ 
200	 & 	354.925690	 & 	48.608368	 & 	1.3 $\pm$ 0.4	 & 	86 $\pm$ 9 \\ 
\multicolumn{5}{c}{}\\ 
201	 & 	354.931885	 & 	48.662445	 & 	0.6 $\pm$ 0.3	 & 	52 $\pm$ 11 \\ 
202	 & 	354.943237	 & 	48.702251	 & 	1.4 $\pm$ 0.4	 & 	31 $\pm$ 8 \\ 
203	 & 	354.951843	 & 	48.617947	 & 	1.8 $\pm$ 0.7	 & 	28 $\pm$ 11 \\ 
204	 & 	354.954773	 & 	48.623917	 & 	1.3 $\pm$ 0.4	 & 	39 $\pm$ 8 \\ 
205	 & 	354.966309	 & 	48.636173	 & 	0.7 $\pm$ 0.2	 & 	33 $\pm$ 7 \\ 
206	 & 	354.967590	 & 	48.639374	 & 	0.7 $\pm$ 0.1	 & 	58 $\pm$ 5 \\ 
207	 & 	354.975006	 & 	48.661385	 & 	0.7 $\pm$ 0.1	 & 	48 $\pm$ 5 \\ \hline
\end{tabular}
\end{minipage}
\end{table}

\begin{figure}[h!]
\centering
\resizebox{9.8cm}{9.5cm}{\includegraphics{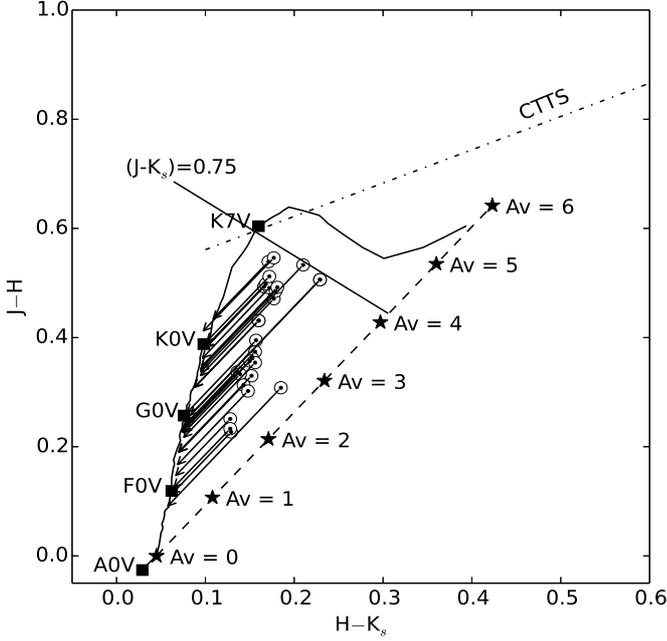}}
\caption{($J-H$) vs. ($H-K_{s}$) CC diagram drawn for stars (with $A_{V}\geq1$) from the region containing Gal 110-13 to illustrate the method. The solid curve represents locations of  unreddened main sequence stars. The reddening vector for an A0V type star drawn  parallel to the \citet{1985ApJ...288..618R} interstellar reddening vector is shown by the dashed line. The locations of the main sequence stars of different spectral types are marked  with square symbols. The region to the right of the reddening vector is known as the near infrared excess region and corresponds to the location of pre-main sequence sources. The dash-dot-dash line represents the loci of unreddened CTTSs \citep{1997AJ....114..288M}. The open circles represent the observed colours, and the arrows are drawn from the observed to the final colours obtained by the method for each star.}\label{fig:CC}
\end{figure} 

\section{DISCUSSION}\label{dis}

\begin{figure}[h!]
\centering
\resizebox{10.5cm}{9.2cm}{\includegraphics{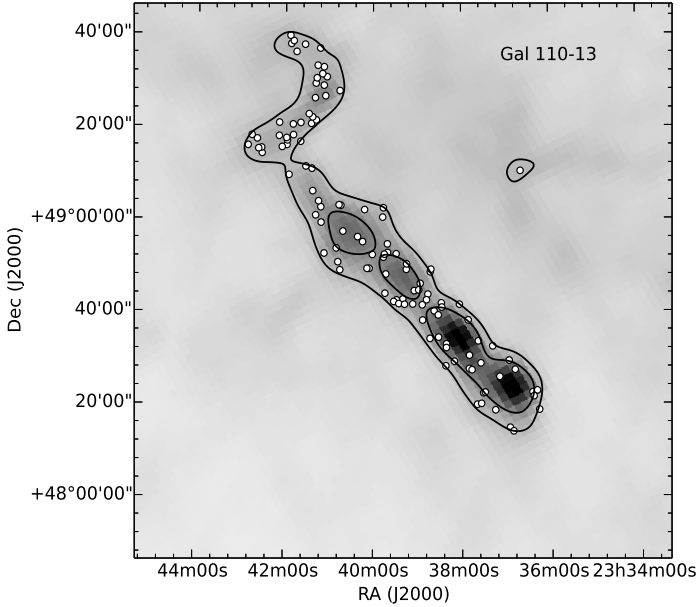}}
\caption{$2\degree\times2\degree$ Planck 857 GHz image of Gal 110-13. The circles represent the positions of those sources that are plotted (closed circles) in the A$_{V}$ vs. distance plot (Fig. \ref{fig:dist}).}\label{fig:gal110img}
\end{figure} 
\begin{figure}[h!]
\centering
\resizebox{9.5cm}{7.4cm}{\includegraphics{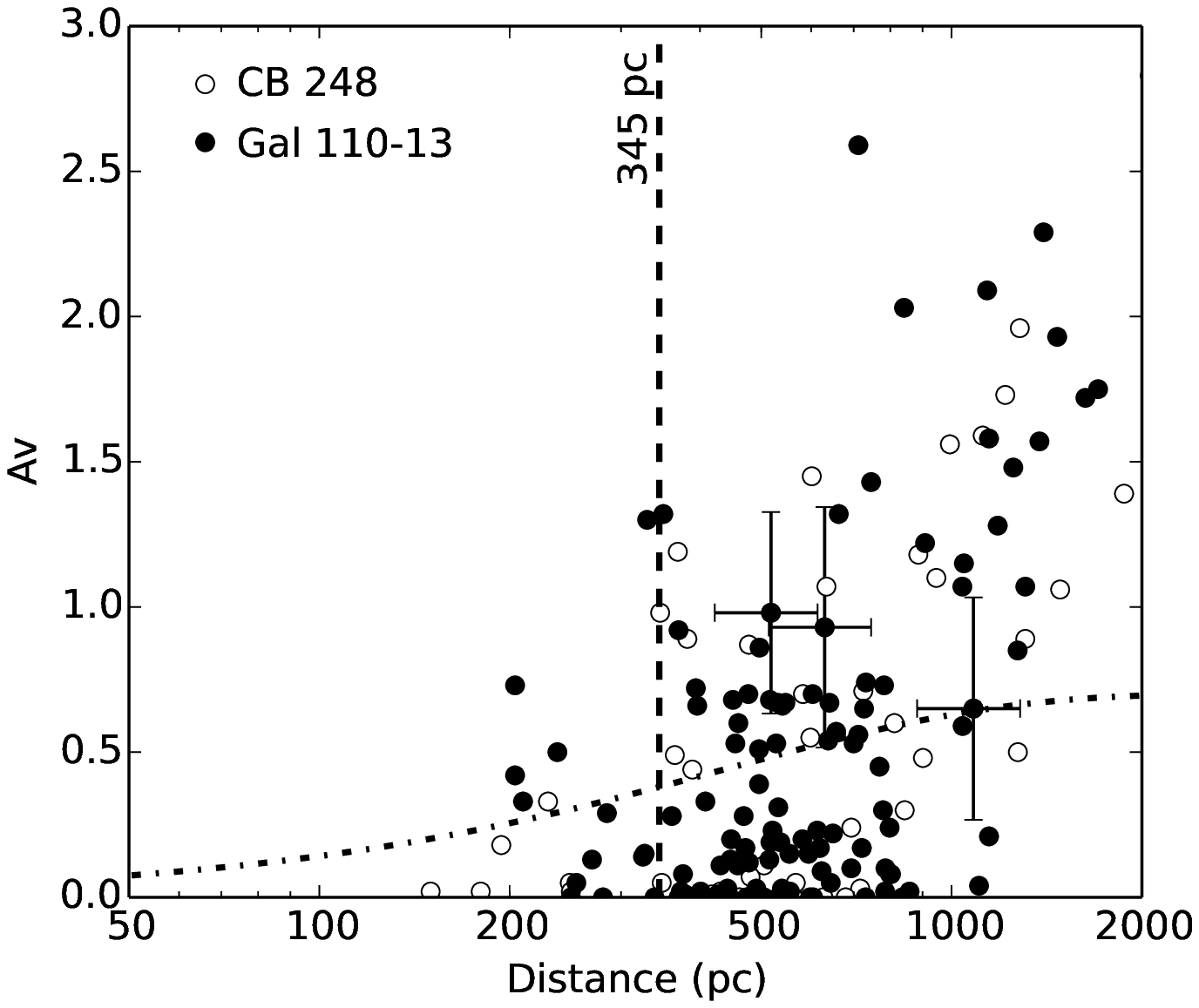}}
\caption{A$_{V}$ vs. distance plot for the regions containing Gal 110-13 (closed circles) and CB 248 (open circles). The dashed vertical line is drawn at 345 pc inferred from the procedure described in \citet{2010A&A...509A..44M}. The dash-dotted curve represents  the increase in the extinction towards the Galactic latitude of $b=-13\degree$ as a function of distance produced from the expressions given by \citet{1980ApJS...44...73B}. Typical error bars are shown on a few data points.} \label{fig:dist}
\end{figure} 

Unpolarized starlight when propagates through interstellar dust grains that are elongated and somehow partially aligned by the interstellar magnetic field, becomes linearly polarized. It appears that the grains are aligned with their shortest axes parallel to the magnetic field direction \citep{1951ApJ...114..206D, 2007JQSRT.106..225L, 2014MNRAS.438..680H, 2014A&A...569L...1A}. The polarization is produced because of the preferential extinction of one linear polarization mode relative to the other. The polarization measured this way gets contributions from all the dust grains that are present in the pencil beam (since a star is a point source) along a line of sight. To obtain the polarization due to the dust grains that are present in the cloud, it is essential to subtract the polarization due to those that are present foreground to the cloud. But to subtract the foreground contribution, the distance of the cloud should be known beforehand. 
  
\subsection{The distance}\label{sec:distance}


\begin{figure}[h!]
\centering
\resizebox{10.5cm}{9.2cm}{\includegraphics{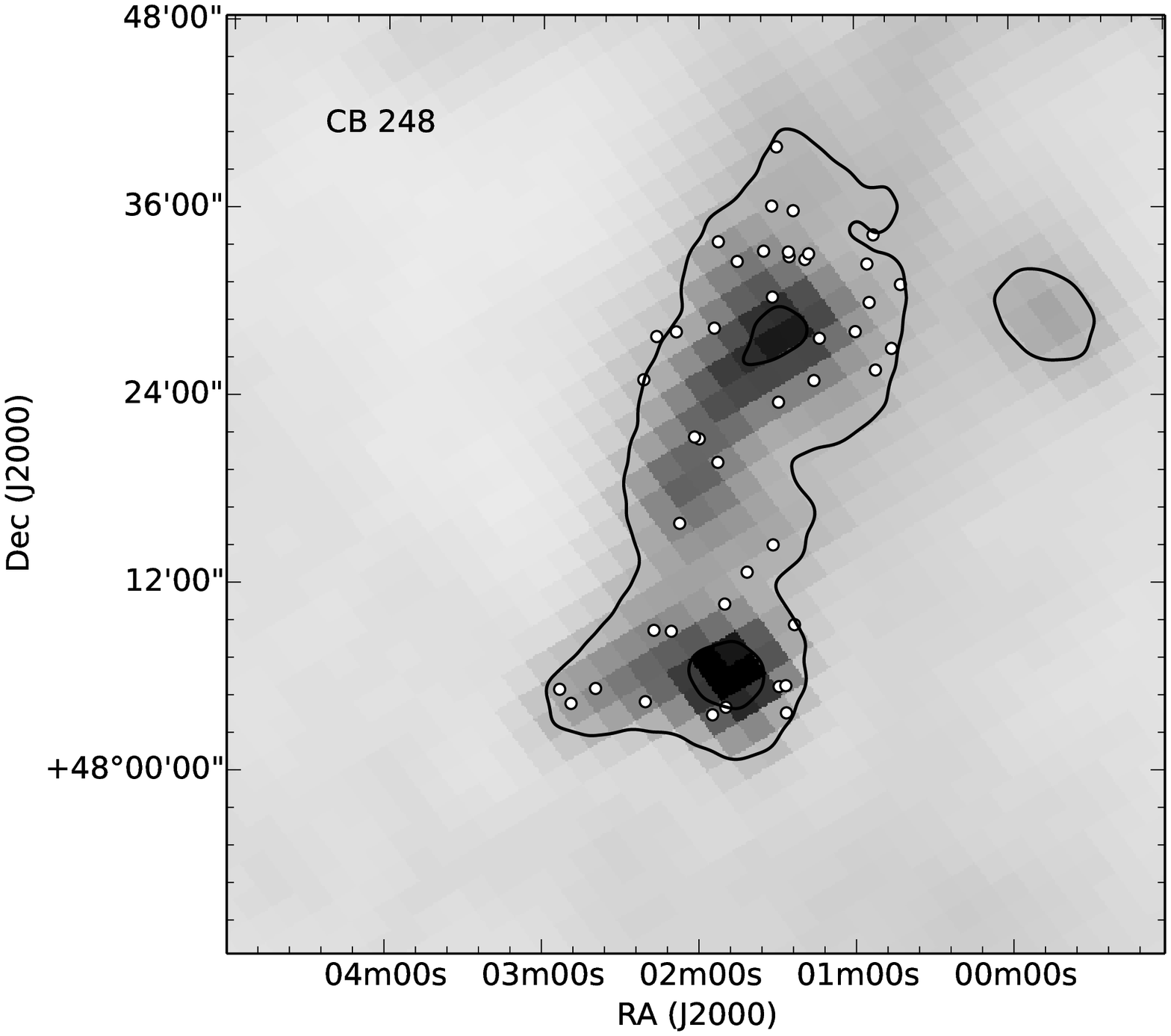}}
\caption{$1\degree\times1\degree$ Planck 857 GHz image of CB 248. The circles represent the positions of those sources that are shown in the A$_{V}$ vs. distance plot (open circles in Fig. \ref{fig:dist}).}\label{fig:cb248img}
\end{figure} 
\begin{figure}[h!]
\centering
\resizebox{8.5cm}{18.5cm}{\includegraphics{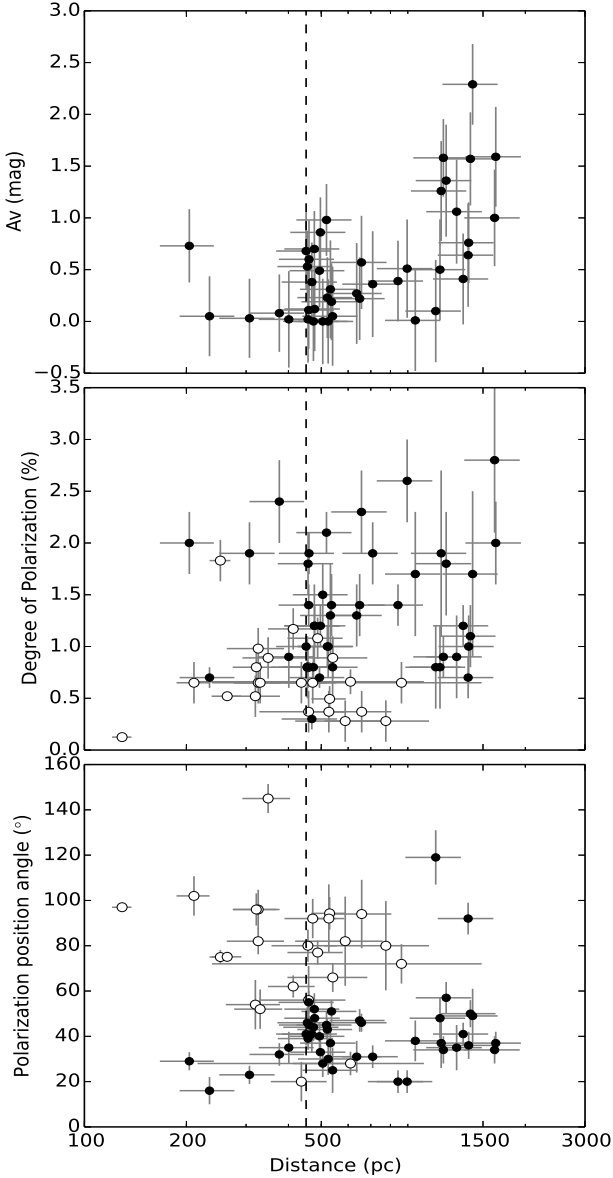}}
\caption{Extinction and distance of 43 stars estimated using the method described in section \ref{sec:distance} (top panel). Degree of polarization (\%) versus distance and position angle versus distance plots for these stars are shown in the middle and bottom panels, respectively. The broken line is drawn at a distance of 450 pc where the first abrupt rise in the value of extinction was seen. The open circles show the polarization results obtained from the \citet{2000AJ....119..923H} catalog of stars selected from a region of 20$\degree\times$20${\degree}$ field containing Gal 110-13 and Gal 96-15 clouds. The distance to these stars are calculated using the new Hipparcos parallax measurements given by \citet{2007A&A...474..653V}.}\label{fig:Av_pol_dis}
\end{figure}

The distance to Gal 110-13 was determined by \citet{1969AJ.....74.1021A} by estimating spectroscopic parallaxes of two B-type stars, HD 222142 (source illuminating the reflection nebula, vdB 158) and HD 222086. They derived a distance of 440$\pm$100 pc to the cloud. However, based on the Hipparcos parallax measurements of HD 222086 \citep{2007A&A...474..653V}, the distance to the star is $\sim640\pm380$ pc. No other distance estimates are available for Gal 110-13 in the literature. We made an attempt to determine the distance to Gal 110-13 using homogeneous $JHK_{s}$ photometric data produced by the Two Micron All Sky Survey \citep[2MASS, ][]{2003yCat.2246....0C}. This is based on a method in which spectral classification of stars projected on the fields containing the clouds are made into main sequence and giants using the $J-H$ and $H-K_{s}$ colors \citep{2010A&A...509A..44M}. In this technique, first the observed $J-H$ and $H-K_{s}$ colors of the stars with ($J-K_{s}$)$\leq0.75$ \footnote{The loci of main sequence stars and the classical T-Tauri stars \citep[CTTS, ][]{1997AJ....114..288M} intercept at ($J-K_{s}$)$\approx0.75$. Therefore the condition ($J-K_{s}$)$\leq0.75$ allows us to exclude both stars later than K7V and CTTS from the analysis.} and photometric errors in $JHK_{s}$ $\leq0.03$ magnitude are dereddened simultaneously using trial values of A$_{V}$  and a normal interstellar extinction law \citep{1985ApJ...288..618R}. The best fit of the dereddened colors to those intrinsic  colors giving a minimum value of $\chi^{2}$ then yielded the corresponding spectral type and A$_{V}$ for the star. The main sequence stars, thus classified, are plotted in an A$_{V}$ versus distance diagram to bracket the cloud distance \citep[e.g.,][]{2010A&A...509A..44M, 2011A&A...536A..99M, 2013A&A...556A..65E, 2013MNRAS.432.1502S}. The pictorial description of the method, given in detail in \citet{2010A&A...509A..44M}, is presented in the near infrared color-color diagram for the stars chosen from the region containing the cloud Gal 110-13 (Fig. \ref{fig:CC}). The extinction vector of an A0V star for increasing values of A$_{V}$ and a normal interstellar extinction law is shown. Extinction values estimated for A0V stars would set the upper limit, and as we move toward more late-type stars, the extinction values traced would decrease in our method.

Any erroneous classifications of giants as dwarfs could result in underestimating their distance, which could lead to uncertainty in any estimation of the distance to the cloud. This problem was resolved by dividing the field containing a cloud into subfields. While the rise in the extinction due to the presence of a cloud should occur almost at the same distance in all the fields, if the whole cloud is  located at same distance, the incorrectly classified stars in the subfields would show high extinction not at same but at random  distances. In case of clouds that have smaller angular sizes, other spatially associated clouds that show similar radial velocities could be selected. Here the assumption is that the clouds that are spatially associated and located at similar distances, have similar velocities.

In Fig. \ref{fig:gal110img}, we show the stars that are lying within the cloud boundary, identified based on the Planck 857 GHz flux, and that are used for estimating the distance of Gal 110-13.  The A$_{V}-$distance plot of these stars are shown in Fig. \ref{fig:dist}, along with the change in the extinction as a function of distance toward the Galactic latitude of $b=-13\degree$ produced using the expressions given by \citet{1980ApJS...44...73B}. An abrupt increase in the values of extinction, values which are significantly above those expected from the expressions by \citet{1980ApJS...44...73B}, is noticed for sources lying beyond $\sim350$ pc. A number of sources lying close to $\sim200$ pc also show a slight increase in the extinction value. 

Based on the 21-cm line observations, \citet{1990RMxAA..21..269C} found three main components having LSR-velocities (V$_{LSR}$) of about 0 to $+$4 km~s$^{-1}$, $-12$ to $-3$ km~s$^{-1}$ and $-30$ to $-20$ km~s$^{-1}$ in the spectra of the region $88\degree\leq l\leq 106\degree$ and $-26\degree\leq b\leq-10\degree$. On account of their spatial distribution, \citet{1990RMxAA..21..269C} suggested that all the three components are related to the region and explained in terms of expansion of a shell of neutral and molecular gas centered at $l=99\degree$ and $b=-12\degree.5$. The radial velocity of 10 Lac (nearest O-type star) is found to be $-9.7$ km~s$^{-1}$ \citep{2009PASP..121.1045K}. The V$_{LSR}$ velocity estimated for Gal 110-13 using $^{12}$CO (J$=$1-0) is found to be $-7.7$ km~s$^{-1}$ \citep{1992ApJ...397..174O}. We searched for additional clouds in the vicinity of Gal 110-13 that show radial velocities similar to that of Gal 110-13. We found one cloud, CB 248, in the catalog produced by \citet{1988ApJS...68..257C}. The V$_{LSR}$ velocity of CB 248 lying $\sim4\degree$ to the east of Gal 110-13 is found to be $-9.5$ km~s$^{-1}$ \citep{1988ApJS...68..257C}. We estimated the distance to this cloud also because of its proximity to Gal 110-13 and of the similar radial velocity. The stars chosen from within the cloud boundary, again identified based on the Planck 857 GHz flux, are shown in Fig. \ref{fig:cb248img}. The A$_{V}$ and the distance values estimated for the selected stars are plotted in Fig. \ref{fig:dist}. A sudden increase in A$_{V}$ values compared to those expected from the expressions given by \citet{1980ApJS...44...73B} is found to occur at $\sim350$ pc and beyond, similar to the result we obtained toward Gal 110-13 region. The vertical dashed line in $A_{V}$ vs. $d$ plot, used to mark the distance of the clouds, is drawn at  $345\pm55$ pc deduced from the procedure described below. We first grouped the stars into distance bins of $bin~width = 0.18\times distance$. The separation between the centers of each bin is kept at half of the bin width. The mean values of the distances and the A$_{V}$ of the stars in each bin were calculated by taking 1000 pc as the initial point and proceeded toward smaller distances. This was done because of the poor number statistics toward shorter distances. The mean distance of the stars in the bin at which a significant drop in the mean of the extinction occurred was taken as the distance to the cloud. The average of the uncertainty in the distances of the stars in that bin was taken as the final uncertainty in distance determined by us for the cloud.

Of the 207 stars for which we have polarization measurements, 43 of them were classified as main sequence stars using the method described above. These sources are those that satisfied the conditions of ($J-K_{s}$)$\leq0.75$ and photometric errors in $JHK_{s}$ $\leq0.03$ magnitude. In Fig. \ref{fig:Av_pol_dis}, the upper panel shows the derived extinction as a function of their distance. The plot shows a rapid increase in the value of extinction close to a distance of $450$ pc (broken line). This slightly disagrees with the distance obtained using extinction of stars from a larger area of Gal 110-13. In the middle and the lower panels, we present the variations in P\% and $\theta_{P}$ as a function of distance. Normally, the degree of polarization, similar to the extinction due to the dust grains, rises with the increase in the column of dust grains along the pencil beam of a star. When the pencil beam of the star passes through a dust cloud, the P\% tends to show a rapid increase in the value if there is no significant change in the direction of the alignment of the dust grains along the line-of-sight. Therefore by combining the polarization and distance information of stars projected on a cloud, the distance to the cloud can be estimated \citep[e.g., ][]{2008A&A...486L..13A}. The measured values of P\% show an increase from $\sim0.5\%$ to $\sim2\%$ at around 450 pc similar to the rise seen in the extinction values. Because the values of P\%  shown in Fig. \ref{fig:Av_pol_dis} are actual measurements, the rise observed in P\% at around 450 pc seems to be more genuine. In that case, the presence of a number of stars with relatively high values of extinction at distances close to 350 pc seen in Fig. \ref{fig:dist} could either be due to wrongly classifying of them as main sequence stars, or that there are additional dust components between Gal 110-13 and us. No further information such as parallax or spectral type is available for these sources in the literature. Employing the same method as used in this work, \citet{2013MNRAS.432.1502S} estimated a distance of $360\pm65$ pc to another cometary globule, Gal 96-15, located at 14$\degree$ to the west of Gal 110-13. The absorbing material at around 350 pc is being consistently detected in the A$_{V}$ versus distance diagrams of both Gal 110-13 and Gal 96-15, suggesting that the presence of this absorbing layer could possibly be real.

To understand the global distribution of interstellar material toward the region containing Gal 110-13 and Gal 96-15, we searched for sources within a region with a 20${\degree}\times$20${\degree}$ field that has both polarization and parallax measurements (\citet{2000AJ....119..923H} and  \citet{2007A&A...474..653V}, respectively). The polarization results are shown in Fig. \ref{fig:Av_pol_dis} (middle and lower panels) and polarization vectors are drawn in Fig. \ref{fig:globalMF}. The length and the orientation of the vectors drawn in Fig. \ref{fig:globalMF} correspond to the P\% and the position angle measured from the galactic north (increasing eastward), respectively. The P\% values from the \citet{2000AJ....119..923H} catalog show an increase at around 200 pc by $\sim0.5\%$ and then a possible increase from $\sim0.5\%$ to $\sim1\%$ at around 320 pc giving evidence of at least two additional dust components lying toward the direction of Gal 110-13. It is quite intriguing that the P\% values show a decreasing trend after 450 pc as against what is expected. The contributions from these foreground absorbing material could be the most likely cause of a few stars in Fig. \ref{fig:dist} showing relatively high values of extinction at distances less than 450 pc (the distance at which the P\% values show a sudden jump in Fig. \ref{fig:Av_pol_dis}).

The B-type star, HD 222142 that illuminates the reflection nebula vdB 158, is found to share a common proper motion with the Lac OB1 association \citep{2007ApJ...657..884L}. This suggests that there could be a possible link between Gal 110-13 and the Lac OB1. The distance estimated to Lac OB1 (which includes the distance to the association as a whole and to the individual subgroups) ranges from $\sim350$ pc to $\sim600$ pc \citep{2009PASP..121.1045K}. Using the available $uvby\beta$ photometry of the stars earlier than A0 type from a 20${\degree}\times$20${\degree}$ field (similar to what is considered by us to obtain polarimetric results from the \citet{2000AJ....119..923H} catalog), \citet{2009PASP..121.1045K} estimate an average distance of $520\pm20$ pc to a group of 12 stars that are often identified as members of the Lac OB 1 association. However, the photometric distance to the most massive member of Lac OB1 association, 10 Lac, is found to be 715$^{+107} _{-92}$ pc \citep{2009PASP..121.1045K}. But the distance of 10 Lac estimated using new Hipparcos parallax from \citet{2007A&A...474..653V} puts the star at 529$^{+70} _{-50}$ pc which is in good agreement with the average distance estimated for the Lac OB 1 association. 

We note that given the galactic latitude of 10 Lac $b=-16.98\degree$, a distance of $\sim700$ pc would place this star about 200 pc away from the galactic plane. This is about a factor of four higher than the typical scale height of star clusters of age $\lesssim10$ Myr \citep{2000AJ....120..314R, 2014MNRAS.444..290B}. The color excess versus distance plot presented by \citet{2009PASP..121.1045K} clearly shows the presence of absorbing material even at $\sim150$ pc, giving additional evidence that there is foreground absorbing material toward the direction of Gal 110-13. Based on optical photometry obtained using PanSTARRS-1, \citet{2014ApJ...786...29S} estimate a distance of $510\pm51$ pc to Gal 96-15. They also find evidence of material distributed at $\sim300$ pc. In a more recent study based on non-LTE analysis of high quality data, \citet{2014A&A...566A...7N} estimate fundamental parameters of 26 early B-type stars in OB associations. Of these, they estimated a distance of 398$\pm$26 pc to HD 216916 (also called EN Lac, B1.5IV) which is considered to be a member of  the Lac OB1 association. 

Although the distance to the Lac OB1 association, to 10 Lac and to the surrounding clouds is still uncertain, we assign 450$\pm$80 pc as the most probable distance to Gal 110-13 based on the results obtained from our polarimetric measurements. The uncertainty in the distance estimated using the 2MASS photometry, described above, is found to be typically be about 18\% \citep[][]{2010A&A...509A..44M}. Additionally, one of the stars, HD 221515, located within an angular separation of $\lesssim1\degree$ from Gal 110-13 and located at a distance of $\sim450$ pc, is found to show polarization angle similar to the average value (45$\degree$) of the measured position angles of the stars projected on the cloud. This gives an additional reason to believe that Gal 110-13 is located roughly at or closer than $\sim450$ pc.

\subsection{The foreground interstellar polarization subtraction}

\begin{center}
 \begin{table}[h!]
	\caption{Polarization values of the seven foreground stars.}\label{table:4}
	\begin{minipage}{\textwidth}  
 \begin{tabular}{l l l l l} \hline
 ID & Source & P $\pm$ $\epsilon_{\mathrm{P}}$ & $\theta$ $\pm$ ${\epsilon_{\theta}}$ & Distance$^{\ddagger}$\\ 
    & identification		&    $(\%)$                  &   $(\degree)$      & (pc) \\ \hline
1   & HD 222104 		& 0.01 $\pm$ 0.06 &  68 $\pm$ 19 & 91 \\
2   & HD 222515 		& 0.37 $\pm$ 0.12 &  94 $\pm$  7 & 173 \\
3   & HD 222590 		& 0.09 $\pm$ 0.08 &  97 $\pm$ 13 & 178 \\
4   & BD+47$\degree$4243& 0.39 $\pm$ 0.10 &  60 $\pm$  6 & 191 \\
5   & BD+47$\degree$4206& 0.15 $\pm$ 0.10 &  59 $\pm$  9 & 242 \\
6   & HD 221775 		& 0.68 $\pm$ 0.09 &  56 $\pm$  3 & 281 \\
7   & HD 221515 		& 0.33 $\pm$ 0.06 &  42 $\pm$  4 & 448 \\ \hline
 \end{tabular}
 
 $^{\ddagger}$Distances are estimated using the Hipparcos parallax meas-\\urements taken from \cite{2007A&A...474..653V}.
 \end{minipage}
 \end{table}
 \end{center}
\begin{figure}
\centering
\resizebox{10.5cm}{8.5cm}{\includegraphics{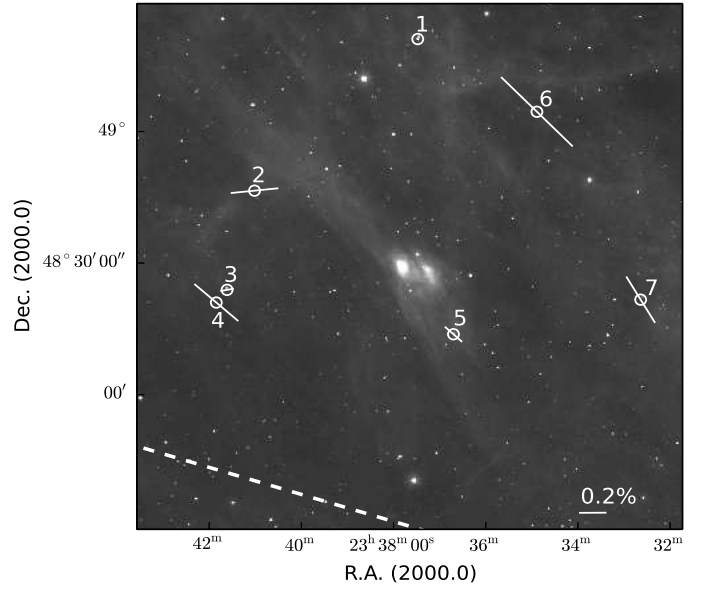}}
\caption{Seven stars having parallax measurements by the Hipparcos satellite within a circular region of 2$\degree$ diameter around Gal 110-13 are identified in the WISE 12$\mu$m image. The polarization vectors are drawn so that the length corresponds to the degree of polarization, and the orientation corresponds to the position angle measured from the north toward the east. A 0.2\% vector is drawn for reference. The broken line shows the orientation of the Galactic plane. Five of these seven stars are used to subtract the foreground interstellar polarization from the observed values.}\label{fig:foreground_img}
\end{figure}
\begin{figure}
\centering
\resizebox{8.5cm}{13.5cm}{\includegraphics{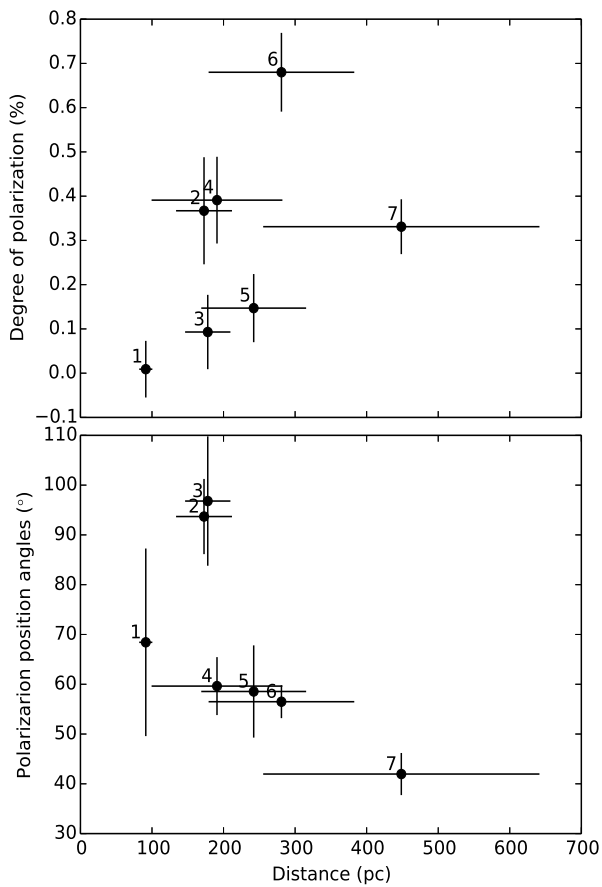}}
\caption{{\bf Upper panel:} Degree of polarization versus distance of the seven stars found within $2\degree$ diameter of Gal 110-13. The distances are calculated from the Hipparcos parallax measurements taken from \citet{2007A&A...474..653V}. {\bf Lower panel:} Polarization position angle versus distance for the same stars. The five stars lying foreground to Gal 110-13 are used to subtract the foreground interstellar polarization from the measured values.}\label{fig:fore_D_P}
\end{figure}
\begin{figure}
\centering
\resizebox{10.0cm}{9.0cm}{\includegraphics{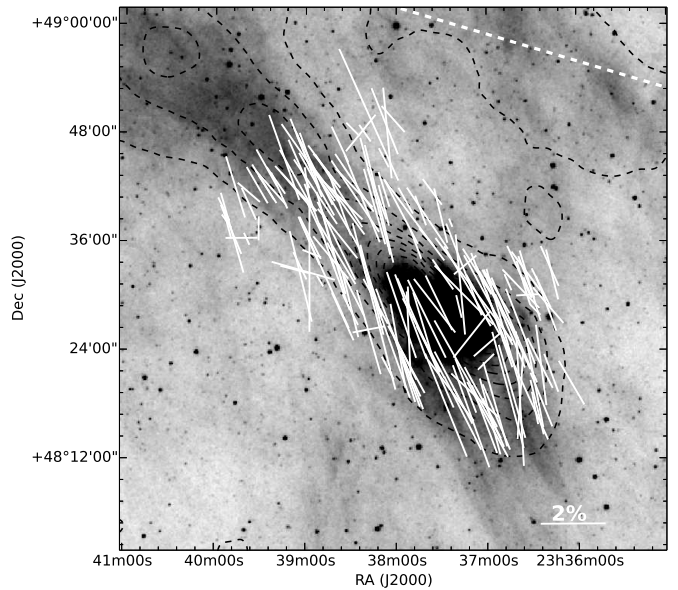}}
\caption{Polarization vectors are over plotted on the WISE 12$\mu$m image of Gal 110-13. The IRAS 100$\mu$m contours are also overlayed on the image. The contours start from 6.0 MJy/sr to 31.2 MJy/sr in increments of 2.8 MJy/sr. The length of the vectors corresponds to the degree of polarization, and orientation corresponds to the position angle measured from the north and increasing toward the east. A vector corresponding to 2\% polarization is shown for reference. The dashed line in white shows the orientation of the Galactic plane.}\label{fig:vectors}
\end{figure} 

To remove the foreground interstellar polarization from the observed polarization values, we made a search around Gal 110-13 within a circular region of 2$\degree$ diameter for stars having parallax measurements made by the Hipparcos satellite in \citet{2007A&A...474..653V}. We excluded those that are identified as emission line sources, in binary or multiple systems or peculiar sources in the SIMBAD. Stars with the ratio of the parallax measurements and the error in parallax $\ge$ 2 were chosen. The distance of these stars range from $\sim90$ to $\sim450$ pc. We made polarization observations of these seven stars in the R-band using AIMPOL. In Table \ref{table:4}, we present the results ordered according to their increasing distance from the Sun. The positions of these stars with respect to Gal 110-13 are shown in the $2\degree \times 2\degree$ WISE 12$\mu$m image in Fig. \ref{fig:foreground_img}.

The P$\%$ and $\theta_{P}$ of the seven stars with their distances are shown in Fig. \ref{fig:fore_D_P}. We estimated the weighted mean of P\% and $\theta_{P}$ using five of the seven stars. We excluded the source at $\sim90$ pc because it shows almost zero polarization. We also excluded the source located at $\sim450$ pc because the $\theta_{P}$ value of this source (42$\degree$) is similar to the mean value of the $\theta_{P}$ obtained for the sources projected on Gal 110-13. We suspect that this source lies either behind or very close to Gal 110-13. The weighted mean values of the degree of polarization and the position angles are found to be 0.3$\%$ and 73$\degree$. 

Using these values, we calculated the mean Stokes parameters Q$_{\mathrm{fg}} (= P \cos2\theta)$ and U$_{\mathrm{fg}} (= P \sin2\theta)$ as $-$0.249 and 0.168, respectively. We also calculated the Stokes parameters Q$_{\mathrm{\star}}$ and U$_{\mathrm{\star}}$, for the target sources. Then we estimated the foreground-corrected Stokes parameters Q$_{\mathrm{c}}$ and U$_{\mathrm{c}}$ of the target sources using 
$Q{_{\mathrm{c}}} = Q{_{\mathrm{\star}}} - Q{_{\mathrm{fg}}}$ and $U{_{\mathrm{c}}} = U{_{\mathrm{\star}}} - U{_{\mathrm{fg}}}$.
The corresponding foreground-corrected degree of polarization (P${_{\mathrm{c}}}$) and the position angle ($\theta{_{\mathrm{c}}}$) values of the target stars were estimated using the equations
$P{_{\mathrm{c}}} = \sqrt{{(Q{_{\mathrm{c}}})^2} + {(U{_{\mathrm{c}}})^2}}$ and $\theta{_{\mathrm{c}}} = 0.5 \times \tan{^{-1}}\Big(\frac{U{_{\mathrm{c}}}}{Q{_{\mathrm{c}}}}\Big)$.
After the foreground subtraction, 186 target stars satisfied the condition of P/$\sigma_{\mathrm{P}}$ $\geq$ 2. The polarization vectors of 186 foreground-subtracted stars are overplotted on the WISE 12$\mu$m image as shown in Fig. \ref{fig:vectors}. The IRAS 100$\mu$m intensity contours are also overplotted. The length and the orientation of the vectors correspond to the measured P$\%$ and $\theta_P$ values, respectively. The $\theta_P$ is measured from the north increasing toward the east. A vector with 2\% polarization is shown for reference as is the orientation of the Galactic plane at the galactic latitude of $-13\degree$.

\subsection{The cause of the unusual structure of the cloud}

\begin{figure}
\centering
\resizebox{9.0cm}{15.5cm}{\includegraphics{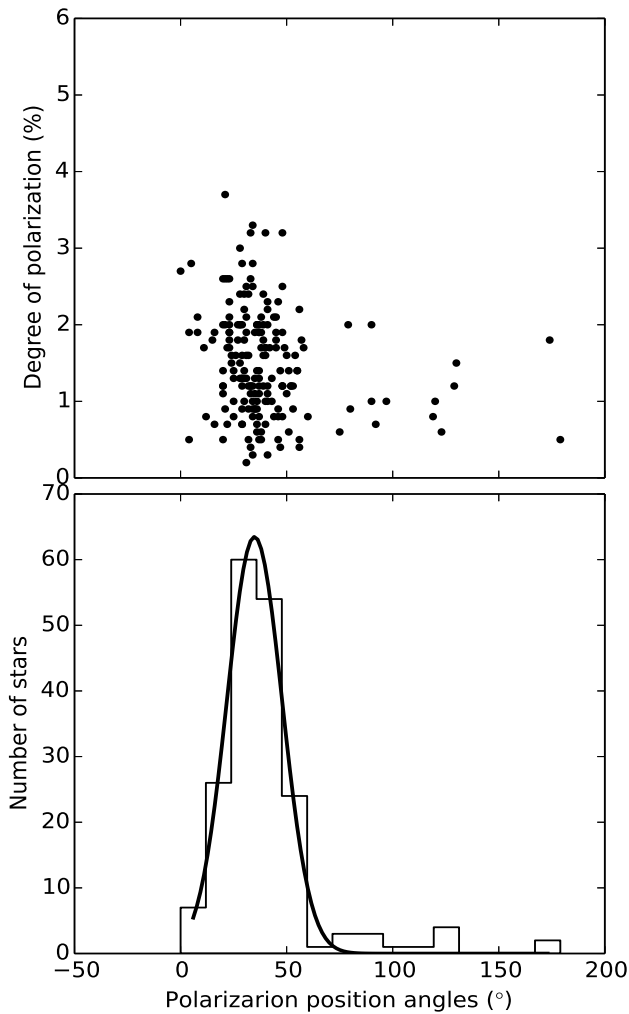}}
\caption{Degree of polarization versus the polarization position angle of stars projected on Gal 110-13. The histogram of the position angles is also presented. These values are obtained after subtracting the foreground interstellar component from the observed values. The solid curve represents a fit to the histogram of position angles.}\label{fig:hist_plot}
\end{figure}
\begin{figure}[h]
\centering
\resizebox{10.0cm}{9.0cm}{\includegraphics{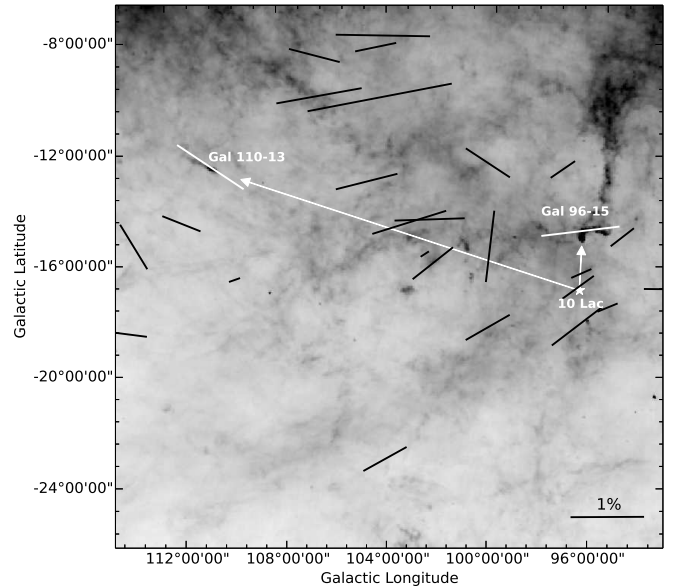}}
\caption{Region containing Gal 96-15, Gal 110-13, and 10 Lac are shown in the IRAS 100$\mu$m image. The ambient magnetic field direction at the location of Gal 96-15 is drawn. The mean magnetic field direction corrected for the foreground interstellar contribution is drawn at the position of Gal 110-13. The arrows show the direction of propagation of ionizing radiation from 10 Lac toward Gal 96-15 and Gal 110-13. The vectors drawn in black are based on the polarization results obtained from the \citet{2000AJ....119..923H} catalog. The length and the orientation of the vectors correspond to the degree of polarization and the position angles measured from the galactic north increasing to the east. A vector corresponding to 1\% polarization is shown for reference.} \label{fig:globalMF}
\end{figure} 

The mean values of P\% after the removal of the foreground interstellar polarization is obtained as 1.5\%. The mean value of $\theta_{P}$ from a Gaussian fit to the data is found to be 40$\degree$ which is considered as the mean direction of the plane of the sky component of magnetic field in Gal 110-13. The plot of P\% vs. $\theta_P$ and the histogram of the $\theta_P$ after correcting for the foreground interstellar contribution are shown in Fig. \ref{fig:hist_plot}. The standard deviation of $\theta_{P}$ values is found to be $\sim11\degree$ implying that the magnetic field lines in Gal 110-13 are ordered relatively well.

Based on far-infrared, HI, and CO data of the region, \citet{1992ApJ...397..174O} proposed cloud-cloud collision scenario as the most preferred mechanism responsible for the formation of Gal 110-13. They suggested that the Gal 110-13 was formed as a result of the interaction between two HI clouds moving across the line of sight and having velocity components of $-8$ and $-6$ km~s$^{-1}$. They also suggested that the southern part compared to the northern part is in an advanced stage that resulted in it being predominantly molecular. Head-on supersonic collision between two identical magnetized clouds was studied using 3D-SPMHD simulations under two special cases of parallel and perpendicular magnetic field orientations with respect to the motion of the colliding clouds \citep{2001A&A...379.1123M}. The magnetic field was amplified and deformed in the shocked layers in both those cases. In the perpendicular magnetic field configuration, owing to the interaction of the clouds, field lines are compressed in the shocked layer forming a magnetic shield that acts like an elastic bumper between the clouds. This prevents the direct contact of the clouds and also their disruption during the interaction \citep{1996ApJ...473..365J}. 

Formation of clumps was noticed only in the case of parallel field configuration. Because star formation is active in Gal 110-13 \citep{1969AJ.....74.1021A, 1992ApJ...397..174O}, and if the cloud was formed as a result of cloud-cloud collision, the initial field configuration would have been parallel to the cloud motion. According to HI observations by \citet{1992ApJ...397..174O}, collided clouds have traveled across the line of sight in the northeast-southwest direction. But the current magnetic field geometry is almost perpendicular to the proposed direction of the interaction of the clouds. Also, the simulations show that in both parallel and perpendicular cases the field distribution after the shock interaction was found to be chaotic especially on the large scales, irrespective of the initial field configuration \citep{2001A&A...379.1123M}. However, the observed magnetic field lines from polarization are found to be uniformly distributed in contrast to the results from the simulations. 

\citet{2007ApJ...657..884L} have considered either a supernova explosion or ionization fronts from a massive star in the vicinity of Gal 110-13 as an alternate mechanism that might have caused its cometary shape. In a supernova scenario, a massive star, which is probably a member of Lac OB1 association to which 10 Lac is considered to be a member, could have exploded as a supernova. The shock waves from such an explosion could have traveled to the location of Gal 110-13 at a speed of hundreds of km~s$^{-1}$ to reach Gal 110-13 and compress the cloud to give its present shape and trigger the star formation. In an ionization-front-induced  formation scenario of Gal 110-13, the ionizing photons from 10 Lac, soon after its birth, might have compressed the cloud, creating a cometary morphology and subsequent star formation \citep{2007ApJ...657..884L}. The spatial separation between Gal 110-13 and 10 Lac is estimated to be $\sim 110$ pc by adopting a distance of 450 pc to Gal 110-13. However, if we consider the most distant distance of 715 pc to 10 Lac estimated by \citet{2009PASP..121.1045K}, and assume that Gal 110-13 also lies at the same distance, then the spatial separation between 10 Lac and Gal 110-13 becomes $\sim180$ pc. As pointed out by \citet{2007ApJ...657..884L}, the ionization front would take about $\sim2-3$ Myr time to travel the distance of $\sim$110-180 pc between 10 Lac and Gal 110-13. This is relatively shorter than the $5.5\pm0.5$ Myr age estimated for 10 Lac by \citet{2011MNRAS.410..190T}.

Based on the radiation-magnetohydrodynamics (R-MHD) simulations of globules \citep{2009MNRAS.398..157H, 2011MNRAS.412.2079M}, it was shown that both radiation-driven implosion and acceleration of clumps by the rocket effect\footnote[2]{Acceleration of the globule away from the ionizing source, caused due to the back reaction of the transonic photo-evaporation flow that leaves the ionization front and moves toward the ionizing source.} tend to align the magnetic field with the direction of propagation of ionizing photons in the globules. The efficiency of this alignment depends on the initial magnetic field strength. While the field reorientation is prevented significantly in a strong magnetic field scenario ($\sim150-180\mu$G), in the case of medium and weak ($\sim50\mu$G) field strengths, the field lines are dragged into alignment with the direction of the ionizing radiation. The structure of a dynamically evolving globule subjected to photo-ionization depends on the initial magnetic field orientation. Simulations with three initial field orientations (perpendicular, parallel, and inclined) with respect to the direction of propagation of the photo-ionizing radiation have been carried out by \citet{2009MNRAS.398..157H}. In the case of strong perpendicular initial magnetic field, the globule evolution is found to be highly anisotropic. The globule becomes flattened in the direction of magnetic field lines. But in the case of strong or weak parallel initial magnetic field, the globule remains cylindrically symmetric throughout its evolution. Also, since the magnetic field opposes the lateral compression of the neutral globule, it becomes  broader with a snubber head. 

The magnetic field strength estimated in Gal 110-13 is found to be $\sim25\mu$G (section \ref{str_fun}) assuming the distance as 450 pc. The region containing Gal 110-13, Gal 96-15 and 10 Lac is shown in Fig. \ref{fig:globalMF}. It is interesting to note that the elongated structure of Gal 110-13 is aligned roughly to the line joining the cloud and the 10 Lac which is $\sim70\degree$ with respect to the galactic north (also considered as the direction of propagation of the ionizing photons from 10 Lac). In Fig. \ref{fig:globalMF}, we also show the mean value of the magnetic field direction in Gal 110-13 drawn at an angle of $\sim60\degree$ with respect to the galactic north. The mean value of the polarization position angles, obtained from the \citet{2000AJ....119..923H} catalog, of the stars located within the 20$^{\degree}\times$20$^{\degree}$ field (as shown in Fig. \ref{fig:globalMF}) is found to be $\sim100\degree$ with respect to the galactic north. This is considered to be the direction of the initial magnetic field prior to the ionization of the region by the 10 Lac. 

The magnetic field geometry of Gal 96-15 was mapped by  \citet{2013MNRAS.432.1502S}. The initial magnetic field orientation toward Gal 96-15 prior to the cloud being hit by the ionizing radiation from the 10 Lac was inferred as $\sim95\degree$  \citep[][]{2013MNRAS.432.1502S} with respect to the galactic north. This is consistent with the direction of the initial field direction inferred using a larger sample of stars distributed over a wider area in this study. Thus the initial magnetic field direction is offset by $\sim30\degree$ with respect to the line joining the cloud and 10 Lac. The ionizing photons from 10 Lac might have reoriented the relatively weak $\sim25\mu$G magnetic field and made it aligned with the direction of propagation, thus creating the unusually long cometary shape of the cloud. On the other hand the line joining the head part of Gal 96-15 and 10 Lac makes an angle of $\sim1\degree$ with respect to the galactic north (see Fig. \ref{fig:globalMF}). The initial magnetic field in Gal 96-15 was therefore oriented almost perpendicular to the direction of propagation of the ionizing photons from the 10 Lac. Accordingly, Gal 110-13 and Gal 96-15 could be considered as good examples of the photo-ionization of clouds with the magnetic field oriented parallel and perpendicular to the direction of propagation of the ionizing radiation, respectively. As also seen in simulations, the difference in the orientation of magnetic field direction with respect to the direction of the propagation of the ionizing photons could be the reason for the difference in the physical structure of both the clouds. While Gal 110-13 shows an elongated cloud structure with $\sim1\degree$ in extent, Gal 96-15 shows a comma structure with the tail part curled almost perpendicular to the line joining the cloud and the 10 Lac and almost aligning with the inferred initial magnetic field direction. This suggests that the magnetic field in Gal 96-15 might be stronger than the field strength estimated in Gal 110-13, causing the field to resist its reorientation along the direction of 10 Lac.

\subsection{The structure function}\label{str_fun}

The strength of the magnetic field can be estimated by determining the turbulent angular dispersion in Gal 110-13. \citet{2009ApJ...696..567H} assumed that the net magnetic field is basically a combination of large-scale structured field, B$_\mathrm{0}$(x), and a turbulent component, B$_\mathrm{t}$(x). Initially in this method the angular dispersion function (ADF) or the square root of the structure function is calculated, which is defined as the root mean-squared differences between the polarization angles measured for all pairs of points (\emph{N}(\emph{l})) separated by a distance \emph{l} (see eq 3). The ADF shows the dispersion of the polarization angles as a function of the distance in a specific region. 

Recently, this method has been used as an important statistical tool to infer the relationship between the large-scale structured field and the turbulent component of the magnetic field in molecular clouds \citep[e.g., ][]{2009ApJ...696..567H, 2010ApJ...723..146F, 2012ApJ...751..138S, 2013A&A...556A..65E}.  The expression for the ADF is 
\begin{equation}
\langle\Delta\phi^{2}(l)\rangle^{1/2} =\Bigg \{\frac{1}{N(l)} \sum\limits_{i = 1}^{N(l)} [\phi(x) - \phi(x+l)]^{2} \Bigg \}^{1/2}
\end{equation}
The total measured structure function within the range $\delta < \emph{l} \ll d$, (where $\delta$ and $d$ are the correlation lengths which characterize B$_\mathrm{t}$(x) and B$_\mathrm{0}$(x), respectively), can be estimated using \citep{2009ApJ...696..567H}:
\begin{equation}
\langle\Delta\phi^{2}(l)\rangle_{\mathrm{tot}} \simeq b^{2} + m^{2}l^{2} + \sigma^{2}_{\mathrm{M}}(l)\label{eq:2}
\end{equation} 
where $\langle\Delta\phi^{2}(l)\rangle_{\mathrm{tot}}$ is the total measured dispersion estimated from the data. The quantity $\sigma^{2}_{\mathrm{M}}(l)$ is the measurement uncertainties, which are calculated by taking the mean of the variances on $\Delta\phi(l)$ in each bin. The quantity $b^{2}$ is the constant turbulent contribution, estimated by the intercept of the fit to the data after subtracting $\sigma^{2}_{\mathrm{M}}(l)$. The quantity $m^{2}l^{2}$ is a smoothly increasing contribution with the length $l$ ($m$ shows the slope of this linear behavior). All these quantities are statistically independent of each other.

The ratio of the turbulent component and the large-scale magnetic fields is given by the equation:
\begin{equation}
\frac{\langle B^{2}_\mathrm{t}\rangle^{1/2}}{B_\mathrm{0}} = \frac{b}{\sqrt{2 - b^{2}}}\label{eq:3}
\end{equation}
 
\begin{figure}
\centering
\resizebox{9.7cm}{7.5cm}{\includegraphics{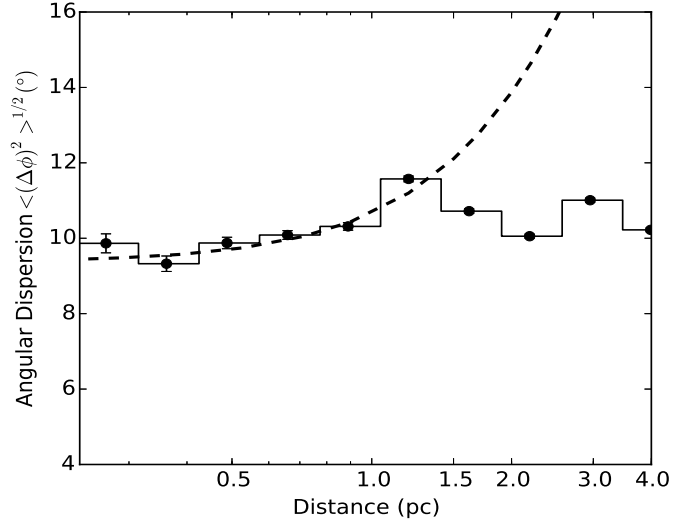}}
\caption{Angular dispersion function (ADF) of the polarization angles, $\langle\Delta\phi^{2}(l)\rangle^{1/2}$ (\degree), with distance (pc) for 169 stars of Gal 110-13. The dashed line denotes the best fit to the data up to $\sim$1 pc distance.}\label{fig:structure_func}
\end{figure}

For the Gal 110-13 region we estimated the ADF and plotted it with the distance in the Fig. \ref{fig:structure_func}. We used the polarization angle of 169 stars to calculate ADF. The measured errors in each bin are very small, which are also overplotted in the Fig. \ref{fig:structure_func}. Each bin denotes the $\sqrt{\langle\Delta\phi^{2}(l)\rangle_{\mathrm{tot}} - \sigma^{2}_{\mathrm{M}}(l)}$ which is the ADF corrected for the measurement uncertainties. Bin widths are taken on logarithmic scale. We used only five points of the ADF in the linear fit of eq. 4. The shortest distance we considered is $\simeq$ 0.1 pc. The net turbulent contribution to the angular dispersion, $b$, is calculated to be 9.5$\degree$ $\pm$ 0.2$\degree$ (0.17 $\pm$ 0.003 rad). Then we estimated the ratio of the turbulent component and the large-scale magnetic fields using eq. 5, which is found to be 0.12 $\pm$ 0.002. The result shows that the turbulent component of magnetic field is very small in comparison to the large-scale structured magnetic field, i.e., $B_\mathrm{t} \ll B_\mathrm{0}$.

The strength of the plane of the sky component of the magnetic field was estimated using the expression \citep{2015ApJ...807....5F},
\begin{equation}
B_\mathrm{pos} = 9.3 \left[ \frac{2~n_{H_{2}}}{cm^{-3}} \right]^{1/2} \left[ \frac{\Delta V}{km s^{-1}} \right] \left[\frac{b}{1^{\circ}} \right]^{-1}~~\mu G \label{equ:B_field}
\end{equation}
in which the scale factor $Q\approx0.5$ is considered based on the results obtained from numerical studies \citep[e.g., ][]{2001ApJ...546..980O}. The above expression was derived by modifying the classical method proposed by \citet{1953ApJ...118..113C} where they suggested that the analysis of the small-scale randomness of magnetic field could be used to estimate the field strength. This method relates the line-of-sight velocity dispersion to the irregular scatter in the polarization position angles under the assumptions that there is a mean field component in the area of interest, that the turbulence responsible for the magnetic field perturbations is isotopic, and that there is equipartition between the turbulent kinetic and magnetic energy \citep{2001ApJ...561..800H}. The volume number density of molecular hydrogen is obtained by estimating the column density of the region probed by our optical polarimetry and from the thickness of the cloud, assuming it to be a cylindrical filament. We used the relation, N(H$_\mathrm{2}$)/$A_\mathrm{v}$ = 9.4 $\times 10^{20}$ cm$^{-2} mag^{-1}$ \citep{1978ApJ...224..132B} for obtaining the column density. Based on the method described in \ref{sec:distance}, the average value of extinction traced by the stars lying behind the cloud (distance $\gtrsim$450 pc) observed in this study is found to be $\sim 0.6$ mag (see Fig. \ref{fig:Av_pol_dis}). The angular diameter of the cloud is found to be $\sim8^{\prime}$. Considering 450 pc as the distance to Gal 110-13, the value of n(H$_\mathrm{2}$) is found to be $\sim175$ cm$^{-3}$. However, if we consider the farthest distance of 10 Lac estimated by \citet{2009PASP..121.1045K} as the distance to Gal 110-13 also, then the value of n(H$_\mathrm{2}$) becomes $\sim110$ cm$^{-3}$.  We obtained the $^{12}$CO line width ($\Delta V$ = 1.4 km s$^{-1}$) toward Gal 110-13 from \cite{1992ApJ...397..174O}. The turbulent contribution to the angular dispersion is found to be $b$ = 9.5$\degree$ $\pm$ 0.2$\degree$. Substituting these values in eq. \ref{equ:B_field}, we obtained $B_\mathrm{pos}$ as $\sim25 \mu$G for a distance of 450 pc and $\sim20\mu$G if the distance is taken as 715 pc. The obtained values of $B_\mathrm{pos}$ should be considered as a rough estimate only mainly due to the large uncertainties involved in the quantities used in their calculations. Apart from the uncertainties in the measurement of line width and in the estimation of $b$ parameter, the dominant source of uncertainty in the above calculation of $B_\mathrm{pos}$ is in the determination of the molecular hydrogen density. 
\label{key}

\subsection{The polarization efficiency}

\begin{figure}
\centering
\resizebox{9.7cm}{15.0cm}{\includegraphics{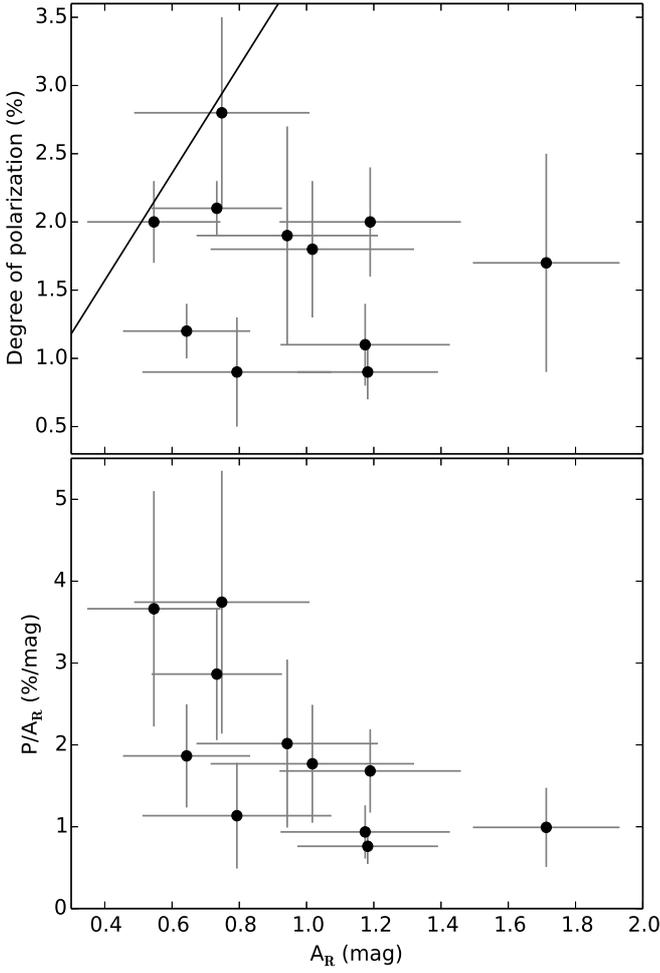}}
\caption{{\bf Upper panel:} P\% versus A$_{R}$ plot of stars for which A$_{R}$/$\sigma$A$_{R}$ $\geq2$. The solid line represents the observational upper limit of the polarization efficiency. {\bf Lower panel:} P/A$_{R}$ versus A$_{R}$ plot.}\label{fig:efficiency}
\end{figure}

The ratio of the degree of polarization to extinction is a measure of the efficiency of polarization produced by the interstellar medium. The efficiency depends on the nature of dust grains and the efficiency with which the dust grains are aligned with the local magnetic field along the line of sight. A theoretical upper limit on the polarization efficiency was calculated by assuming dust grains of infinite cylinders with diameter comparable to the wavelength and the long axes aligned perfectly parallel to one another. 

We determined A$_{V}$ values for 43 stars toward Gal 110-13 using the method mentioned in \ref{sec:distance}. In Fig. \ref{fig:efficiency} (upper panel), we show the P$_{R}$ as a function of A$_{R}$ values obtained using A$_{R}$/A$_{V}$=0.748 \citep{1985ApJ...288..618R}. The 11 stars are plotted with the condition A$_{V}$/$\sigma$(A$_{V})\geq2$. The solid line is drawn for the observational upper limit evaluated for the $\lambda_{\mathrm{eff}}$ = 0.760$\mu$m. The relation between P$_{V}$ and P$_{R}$ is obtained using the empirical formula given by \citet{1975ApJ...196..261S}. Here we used the typical values of the parameter $K=-1.15$, which determines the width of the peak of the curve, and $\lambda_{max}=0.55\mu$m for the calculations, which may depend on observer's line of sight \citep{1992dge..book.....W}. The transformed observational upper limit is found to be P$_{R}$/A$_{R}\approx4\%/$mag. Toward Gal 110-13 region, the polarizing efficiency of dust grains are found to be below the observational limit. In Fig. \ref{fig:efficiency} (lower panel), we present polarization efficiency as a function of extinction. The data are consistent with a trend toward very rapidly decreasing P$_{R}$/A$_{R}$ with the extinction. The polarizing efficiency tends to be higher for stars with low extinction and becomes lower for those with high values of extinction. Thus the population of dust polarizing the background starlight are dominated by those lying in the outer layers of the cloud and the field orientation probed using optical polarization are predominantly of the one on the periphery of the clouds.

\section{Conclusions}\label{con}
We present R-band polarization measurements of 207 stars projected on an unusual, cometary shaped cloud, Gal 110-13. The main goal of the study was to identify the most likely mechanism responsible for its cometary structure and ongoing star formation activity in the cloud. Based on the results from our polarization measurements of stars projected on the cloud and evaluating their distances using 2MASS photometry, we estimated a distance of $450\pm80$ pc to Gal 110-13. The foreground interstellar polarization component was removed from the observed polarization measurements by observing a number of stars located within a circular region of 2$\degree$ diameter and having parallax measurements made by the Hipparcos satellite. The magnetic field thus inferred from the foreground corrected polarization values are found to be parallel to the elongated structure of the cloud. The field lines are found to be distributed uniformly over the cloud. The plane of the sky magnetic field strength estimated using the structure function analysis was found to be $\sim25\mu$G. Based on our polarimetric results, we suggest that the compression of the cloud by the ionization fronts from 10 Lac is the most likely mechanism responsible for creating the cometary shape and the star formation activity in Gal 110-13.

\begin{acknowledgements}
The authors are very grateful to the referee, Prof. Gabriel Franco, for the constructive comments and suggestions, that helped considerably to improve the content of the manuscript. This research made use of the SIMBAD database, operated at the CDS, Strasbourg, France. We also acknowledge the use of NASA's \textit{SkyView} facility (http://skyview.gsfc.nasa.gov) located at NASA Goddard Space Flight Center. This research also made use of APLpy, an open-source plotting package for Python hosted at http://aplpy.github.com. C.W.L was supported by Basic Science Research Program though the National Research Foundation of Korea (NRF) funded by the Ministry of Education, Science, and Technology (NRF-2013R1A1A2A10005125) and also by the global research collaboration of Korea Research Council of Fundamental Science \& Technology (KRCF). NS thanks Suvendu Rakshit (OCA, Nice) for his valuable support and useful discussions.
\end{acknowledgements}

\bibliographystyle{aa}
\bibliography{references}

\label{lastpage}

\end{document}